\documentclass[a4paper,twosides]{article}
\usepackage{CJK,multicol,multirow,graphics,fancyhdr,epstopdf}
\usepackage{amsmath,amsfonts,amssymb,bm,upgreek,mathrsfs,ccmap,mathcomp}
\usepackage[pagewise,switch,columnwise]{lineno}
\usepackage[compress,nospace]{cite}
\usepackage[dvipsnames]{xcolor}
\usepackage{CPL-2023}
\usepackage{makecell}
\usepackage{fontsize}
\usepackage{lineno}
\newcommand{\cplyear}{2024} \newcommand{\cplvol}{41}
\newcommand{\cplno}{x} \newcommand{\cplpagenumber}{111201}
 \newcommand{\cplpage}{\cplpagenumber-\thepage}

\newlength\cmsTabSkip

\begin{document}

\begin{CJK}{GBK}{song}\vspace* {-4mm} \begin{center}
\large\bf{\boldmath{Experimental Road to a Charming Family of Tetraquarks\ldots and Beyond}}
\footnotetext{\hspace*{-5.4mm}$^{*}$Corresponding authors. Email: kai.yi@njnu.edu.cn

\noindent\copyright\,{\cplyear}
\href{http://www.cps-net.org.cn}{Chinese Physical Society} and
\href{http://www.iop.org}{IOP Publishing Ltd}}
\\[5mm]
\normalsize \rm{}Feng Zhu$^{1}$, Gerry Bauer$^{1}$, and Kai Yi $^{1,*}$
\\[3mm]\small\sl $^{1}$School of physics and technology, \\Nanjing Normal University, Nanjing 210046, China
\\[4mm]\normalsize\rm{}(Received xxx; accepted manuscript online xxx)
\end{center}
\end{CJK}
\vskip 1.5mm

\small{\narrower Discovery of the X(3872) meson in 2003 ignited intense interest in exotic (neither $q\bar{q}$ nor $qqq$) hadrons, but a $c\bar{c}$ interpretation of this state was difficult to exclude.
An unequivocal exotic was discovered in the $Z_c(3900)^+$ meson---a {\it charged} charmonium-like state.
A variety of models of exotic structure have been advanced but consensus is elusive. 
The grand lesson from heavy quarkonia was that heavy quarks bring clarity.
Thus, the recently reported triplet of all-charm tetraquark candidates---$X(6600)$, $X(6900)$, and $X(7100)$---decaying to $J/\psi\,J/\psi$ is a great boon,  promising important insights. We review some history of exotics, chronicle the road to prospective 
all-charm tetraquarks, discuss in some detail the divergent modeling of $J/\psi\,J/\psi$  structures, and offer some inferences about them.
These states form a Regge trajectory and appear to be a family of radial excitations. 
A reported, but unexplained, threshold excess could hint at a fourth family member.
We close with a brief look at a step beyond: all-bottom tetraquarks.

\par}\vskip 3mm
\normalsize\noindent{\narrower{PACS: xxxx}}\\
\noindent{\narrower{DOI: \href{http://dx.doi.org/10.1088/0256-307X/\cplvol/\cplno/\cplpagenumber}{10.1088/0256-307X/\cplvol/\cplno/\cplpagenumber}}

\par}\vskip 5mm
\section{Introduction}
\indent

The quark model of hadrons celebrates its diamond jubilee in 2024. Way back in 1964 Gell-Mann~\cite{GellMann} and Zweig~\cite{Zweig} famously proposed  that hadrons are composed of pairs, or triplets, of particles, whimsically  named ``quarks.''
Their idea triumphed in the early 1970s, and it stands at the irrevocable core of the Standard Model of particle physics.
So it was with great excitement that the new millennium opened with the discovery of  the ``$X(3872)$'' meson~\cite{Belle:X3872}---a charmonium-like particle that ushered in a new era of tetra- and penta-quark states.

In the ensuing years there was a veritable flood of exotic candidates~\cite{ali_maiani_polosa_2019}, but controversy persists over the nature of their internal structure, or even if 
some of them might instead  be dynamical artifacts of production thresholds~\cite{Dong:2020nwy-Duplct,Guo:2019twa,Gong:2020bmg}.
A significant handicap was that, as quark states, these candidates contained  light ($u$, $d$, $s$) quarks, and these are notoriously challenging to model because their dynamics is in the non-perturbative regime of QCD.
A boon to model builders would be  exotic hadrons composed  solely of heavy ($c$, $b$) quarks\cite{YI:2013iok}.

This dream was realized in 2020 when the LHCb Collaboration reported a  peak around 6900~MeV in the $J/\psi\,J/\psi$ channel~\cite{LHCb:2020bwg}---a prime candidate for an all-charm tetraquark.
Confirmation came from the ATLAS~\cite{ATLAS:2023bft} and CMS~\cite{CMS:2023owd} Collaborations.
The bounty increased in 2022 when the latter also reported structures around 6600 and 7100 MeV~\cite{CMS:2022pas}.
A heavy triplet offers a qualitative and quantitative bonus over a lone state for model testing, and thus the $J/\psi\,J/\psi$ structures are a singular aid to theorists  unraveling the nature of exotic hadrons.

This article will briefly review the winding road that led to the $X(3872)$,   the experimental campaign towards all-heavy tetraquarks, and finally arriving at a family of all-charm tetraquark candidates.
The current  data on the $J/\psi\,J/\psi$ mass spectrum are reviewed.
We bring coherence to the somewhat divergent  results, and offer some implications, speculations, and perhaps even conclusions.
We close with some  future prospects.

\subsection{Pealing back the layers of matter}
\indent

Since humans have had awareness, we have always asked ourselves where we come from, and how we are constructed.
The Chinese saying ``Tao begets One, One begets Two, Two begets Three, and Three begets all things" reflects humanity's deep curiosity about the fundamental components of our existence---and foreshadows today's vision. 
After a few millennia of such philosophical musings from the ancient Chinese, Greek, and Indian civilizations about the fundamental stuff of our material world, this spirit was finally transformed into a scientific theory of elementary and indestructible entities by John Dalton in 1803~\cite{Dalton}.
%
However, in the early decades of the 20th century it was learned that Dalton's structureless and indestructible atoms had structure (Rutherford in 1911~\cite{Rutherford}) and were being destructed (Cockcroft and Walton in 1932~\cite{CockcroftWalton1932}). 
By mid-century doubts grew that the mounting collection  of elementary particles might not be so elementary  (e.g. Ref.~\cite{FermiYang} for early qualms).

A new and deeper level of structure was posited by Murray Gell-Mann~\cite{GellMann} and George Zweig~\cite{Zweig} in 1964: the quark model.
Inspired by the 1956 model of Shoichi Sakata---where hadrons were composed of elementary $p$, $n$, and $\Lambda^0$~\cite{Sakata:1956}---Gell-Mann and Zweig instead postulated a triplet of quarks, $u$, $d$, and $s$, as constituents.
Mesons are composed of a quark and antiquark ($q\bar{q}$), and baryons of three quarks ($qqq$).
Quarks were odd beasts, they have  fractional electrical charge, and it was perplexing how no one had  stumbled upon fractionally charged matter.
As a measure of the incredulity:   Zweig was unable to publish his paper~\cite{Riordan:1987}!
%
%
It took almost a decade for physical quarks to become orthodoxy. 
The tide turned with deep-inelastic $e$-$p$ scattering at SLAC  in 1969, providing the first direct evidence of quark-like substructure of the proton~\cite{SLAC-ep-parton}.
The 1970s witnessed a prolific age of discovery:  the $J/\psi$ in 1974~\cite{JPsiMIT,JPsiSLAC} and new  world of charm hadron spectroscopy;  and likewise for bottom hadrons starting in 1977~\cite{E288:Upsilon1977}; 
%
particle jets from quarks~\cite{Hanson:jetsSLAC1975};
formulation of the inter-quark force, quantum chromodynamics (QCD)~\cite{Fritzsch:QCD1973}, and its asymptotic freedom~\cite{Gross:AsympFree1973,Politzer:AsympFree1973}.
Later decades witnessed the ever narrowing of gaps, including
tracking down top, the final quark, 
in  1995~\cite{CDF:Top1995,D0:Top1995};
and the last missing ground-state meson, the $B^+_c$ in 1998~\cite{CDF:Bc1998}. 
At the close of the 20th century the quark model of hadrons was utterly triumphant.

Thus it was quite a surprise, only  a few years into the 21st century, when the $X(3872)$ meson was discovered, sparking  
intense interest in  ``exotic'' hadrons, i.e., states that are neither $q\bar{q}$ nor $qqq$.
Despite the deep skepticism over
the quark model prior to the 1970's, the history of exotics predates its acceptance.
In 1949 Fermi and Yang considered $N\bar{N}$ bound states as a model of the pion~\cite{FermiYang}. 
In 1961 the deuteron was put in a dibaryon  multiplet~\cite{Neeman:1961} using the SU(3) symmetry~\cite{Gell-Mann:Symm1962}, and there was some evidence for a $\Lambda p$-state~\cite{Oakes:1963}.
Then, Gell-Mann and Zweig did note in their 1964 papers~\cite{GellMann,Zweig}, that their symmetry relations also permitted tetra- and penta-quark states.

With skepticism over the reality of quarks---versus mere mathematical constructs---it seems doubly ridiculous to entertain exotics.
On the other hand, disbelief in quarks meant one was not chained to its restrictions.
In the mid-1960s enhancements in $KN$ scattering pointed to $+1$ strangeness baryons, implying $qqqq\bar{s}$~\cite{Cool:KN1966}. 
Similarly, $K\bar{K}$ states were suggested to explain a low-mass enhancement in $\bar{p}p {\to} K\bar{K}\pi$~\cite{Astier:KK1967}. And theoretically, duality arguments for baryon-antibaryon
scattering via meson exchanges implied, in quark language, $qq\bar{q}\bar{q}$ systems~\cite{Rosner:2Baryon1968}.

With the embrace of quarks, and  QCD, in the early 1970s, the question of exotics shifted to a dynamical one:
are any manifest in an observationally meaningful way?
QCD opened the door to  new types of exotics: ``hybrids'' with valence gluons, and ``glueballs'' without any quarks.
Constructing a bag model, Jaffe and Johnson  answered positively, but also argued some known $0^{++}$ mesons ($f_0$, $a_0$,. . . ) were better viewed as $qq\bar{q}\bar{q}$ systems~\cite{Jaffe:Exotics1975,Jaffe:I1976,Jaffe:II1976}.
Later, a $K\bar{K}$ state was invoked to explain $\pi\pi {\to}  f_0(980) {\to} K\bar{K}$ data~\cite{Wicklund:KK1980}.\ldots and so it went\ldots

The prospects for exotics ebbed and flowed.
In 1978 the Particle Data Group (PDG) introduced a dibaryon mini-review in its Particle Properties~\cite{ParticleDataGroup:1978}---but dropped it in 1992~\cite{ParticleDataGroup:1992}.
Non-$q\bar{q}$ candidates fared better, with  mini-reviews starting in 1988~\cite{ParticleDataGroup:1988}, and continuing to this day. 
The century closed with some light exotic candidates, but crisp clear examples eluded searchers~\cite{Landsberg:1999Exotic}.
The basic difficulty is that light mesons rapidly broaden as excited, and  potential exotics mix with their ordinary counterparts.
Indeed, the PDG no longer  separates light exotics but subsumes them under ``Scalar mesons below 1 GeV.''
One could conclude that there very likely {\it were} light exotics, but the muddled soup of light mesons left one unsatisfied by the lack of a clear smoking gun.
Experimentalists and theorists alike would continue to slog away trying to unravel light exotics.

As we continue on to the next phase of our exotic story, we summarize in Table~\ref{tab:Detectors} the principal experimental actors that we will encounter.

\begin{table}[tb]
    \centering
    \caption{Summary of the principal players hunting for heavy exotic hadrons.}

    \begin{tabular}{@{}cccccc@{}} 
    \hline
    accelerator&Type&$\sqrt{s}(GeV)$&Runing years&Detectors&location\\
    \hline
    \hline
    Tevatron&$p\bar{p}$&\makecell[c]{1800(Run-I)\\1960(Run-II)}&\makecell[c]{1987-1996(Run-I)\\2000-2011(Run-II)}&CDF, D0&Fermilab\\
    
    KEKB&$e^-e^+$&10.58(8.0/3.5)&1998-2010&Belle&KEK\\
    
    Super-KEKB&$e^-e^+$&10.58(7.0/4.0)&2018-?&Belle II&KEK\\
    
    PEP-II&$e^-e^+$&10.58(9/3.1)&1999-2008&\textsc{BaBar}&SLAC\\
    
    BEPCII&$e^-e^+$&1.84-4.95 &2009-2023&BESIII&IHEP\\
    
    LHC&$pp$&\makecell[c]{ \textgreater 7000(Run-I)\\13000(Run-II)\\13600(Run-III)}&\makecell[c]{2009-2013(Run-I)\\2015-2018(Run-II)\\2022-?(Run-III)}&\makecell[c]{CMS, ATLAS,\\ALICE, LHCb}&CERN\\
    \hline
    \end{tabular}
    \label{tab:Detectors}
\end{table}

\section{Discovery and Establishment of Heavy Exotic Hadrons}
\indent

The new century began with a murky picture of potential light exotics lurking among the standard mesons, but then exotics  were suddenly thrown in the limelight.
At the October 2002 {\sc PANIC} (Osaka) conference the LEPS Collaboration reported  a pentaquark:   four quarks plus one antiquark. 
They saw a peak in
$\Theta^+{\to} n K^+$  at $1540 \pm 10$~MeV\footnote{We adopt the convention, unless specified otherwise, that for experimental results a single uncertainty is statistical only, whereas dual uncertainties are statistical followed by systematic.}~\cite{Muramatsu:ThetaPlus2002}---being a baryon with $+1$ strangeness it was manifestly exotic.
The DIANA Collaboration rapidly followed with a $\Theta^0{\to} p K^0_S$ signal at $1539 \pm 2$~MeV~\cite{DIANA:ThetaZero2003}.
There had been spurious pentaquark claims in the past, but these new results were followed by mounting confirmations in 2003 and beyond~\cite{Trilling-PDG:2006}.

Then on 10 August 2003 Belle caused a stir when they posted to the  arXiv  server their submission to the {\sc 21st Lepton and Photon Symposium} reporting a peak in $J/\psi$$\pi^+\pi^-$  at  $3871.8 \pm 0.7 \pm 0.4$~MeV~\cite{Belle:X3872-LepPhot2003}.
This did not match expectations for standard charmonia---the world of exotics had, maybe, graduated to heavy quarks.

One could cheer 2003 as shaping up to be an {\it annus mirabilis} for exotics.
Unfortunately, as the dust settled, and despite {\it many} confirmations, the $\Theta$ pentaquarks evaporated~\cite{Schumacher:NoPentaQ2005,Trilling-PDG:2006,Wohl:PentaQ2008,Hicks:NoPentaQ2012}.
What about Belle's peak?

\subsection{ Discovery of the $X(3872)$ meson } 
\indent

\begin{figure}[tb] 
    \centering
    \begin{minipage}[b]{0.48\textwidth}
        \centering
        \resizebox{0.96\linewidth}{!}{\includegraphics{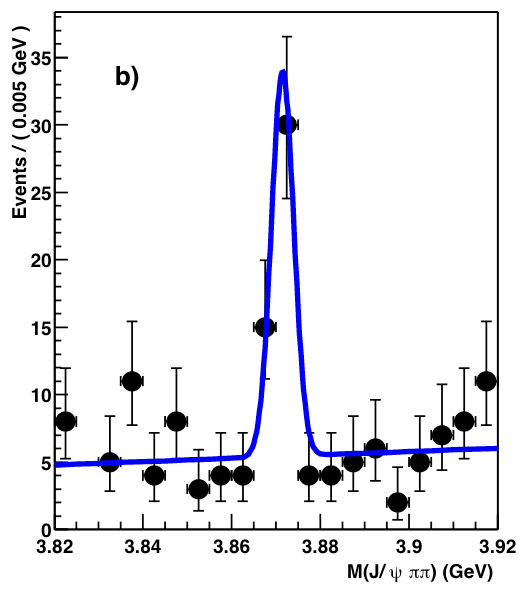}}
        \caption{Belle's $J/\psi \pi^+ \pi^-$ mass distribution showing the $X(3872)$  peak~\cite{Belle:X3872}.}
        \label{fig:Belle3872}
    \end{minipage}
    \hfill
    \begin{minipage}[b]{0.48\textwidth}
        \centering
        \resizebox{0.96\linewidth}{!}{\includegraphics{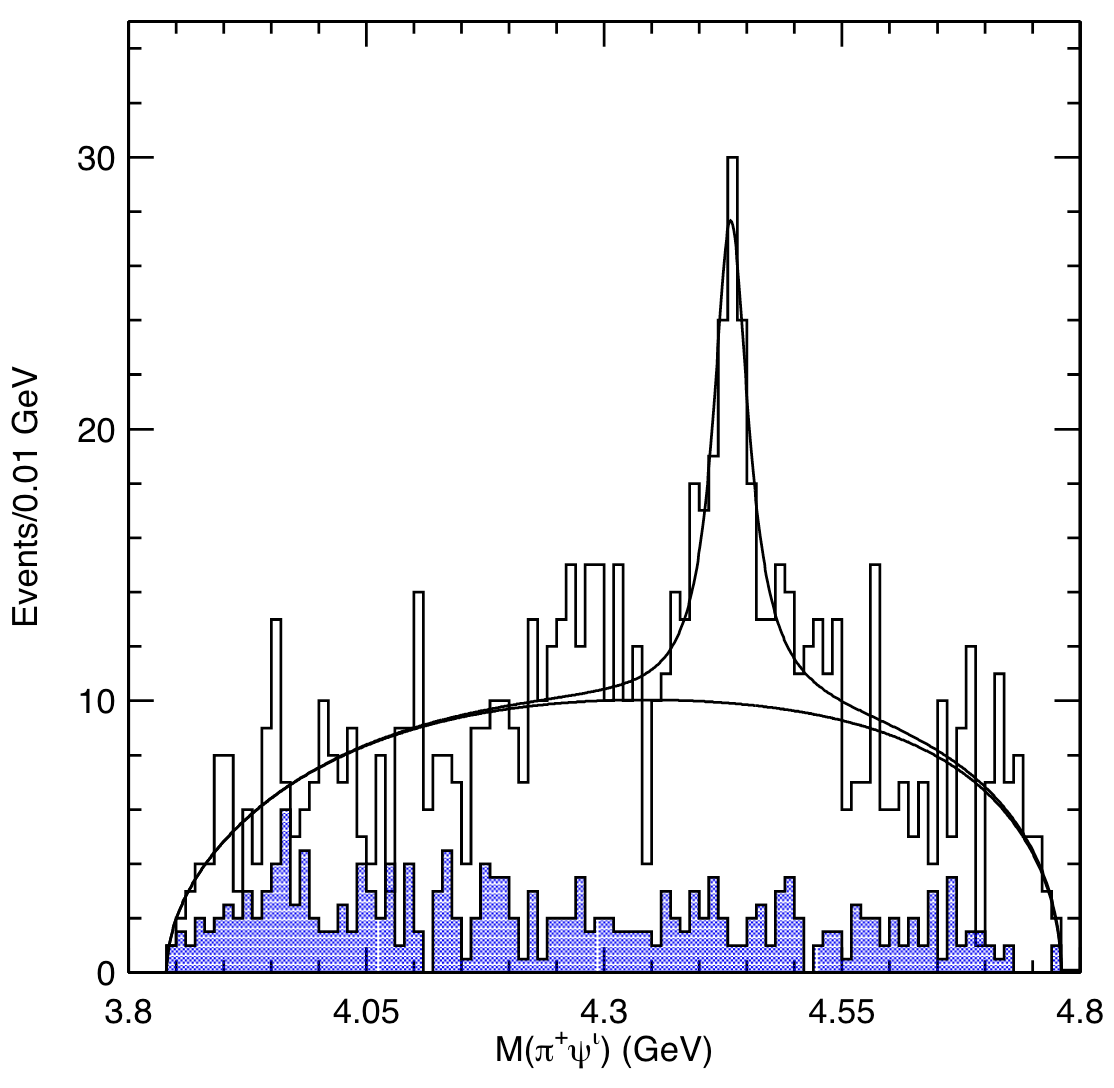}}
        \caption{The $Z_c(4430)^+$ in the $M[\psi(2S) \pi^+]$ distribution from Belle for events in the $M_{bc}-\Delta E$ signal region, where $M_{bc}$ is the beam-constrained mass (defined as $M_{bc} = \sqrt{E_{\text{beam}}^2 - p_B^2}$, $E_{\text{beam}}$ is the cms beam energy, and $p_B$ is the vector sum of the cms momenta of the $B$ decay products)~\cite{Belle:2007hrb}.}
        \label{fig:Z4430}
    \end{minipage}
\end{figure}

In early 2003 a Belle analysis team was optimizing a selection of $B$ decays to $CP$ eigenstates, including  $B \to \psi(2S)K$~\cite{Olsen:2024}.
Fortuitously, the $J/\psi \pi^+\pi^-$ mass window was broad, and a small $3\sigma$-ish bump around 3870~MeV was noticed.
That summer a large dump of new data arrived, and the bump became  an 8.6$\sigma$ peak---the  $X(3872)$\footnote{Some argue a 1993 report by  E705 (FNAL) of a  $J/\psi\pi^+\pi^-$ peak  at  3836~MeV may  have  been  the $X(3872)$:
the PDG separated the  $X(3872)$ and $\psi(3836)$~\cite{ParticleDataGroup:2004}, but later put both under the $X(3872)$~\cite{ParticleDataGroup:2006}.
 E705 fired  300~GeV $\pi^\pm$'s
into a $^7$Li target,
and reported  
$58 \pm 21$ candidates  (local $2.8\sigma$) at $3836 \pm 13$~MeV  (width of  $24 \pm 5$~MeV)~\cite{E705:3836PsiPiPi1993}.
This seems unlikely to be the $X(3872)$.
Aside from  having the wrong mass---it is speculated that maybe the  pion momentum scale was incorrect [but not noticed for the   $\psi(2S)$]---their peak seems too large.
The ratio of the raw  $\psi(3836)$ yield to that of   $\psi(2S)$ was 
 $0.75 \pm   0.34$ ($\sqrt{s} = 24$~GeV), whereas the  raw ratio from CDF's $X(3872)$ confirmation  in $\bar{p}p$ ($\sqrt{s} = 2000$~GeV) was 
$0.100 \pm 0.017$~\cite{CDF:3872-2003}.
(A ratio from LHCb for prompt production  is
$ 0.076 \pm 0.005 \pm 0.009$~\cite{LHCb:3872xSec2021}.)
Also,  CDF's  mass resolution was almost six times better than E705's, and with their resolution CDF would have had difficulty seening an $X(3872)$  in 2003.
The evidence is against an $X(3872)$ interpretation, and with a $2.8\sigma$ local significance, there is little reason not to accept the $\psi(3836)$ as a spurious signal.
}---just in time for August's {\sc Lepton-Photon} symposium.
Belle 
used  152M $B\bar{B}$ events and reported  a peak of $34.4 \pm 6.5$ candidates in the $J/\psi$$\pi^+\pi^-$ channel, with
a mass of $3871.8 \pm 0.7 \pm 0.4$~MeV and a width less than 3.5~MeV (Fig.~\ref{fig:Belle3872})~\cite{Belle:X3872-LepPhot2003,Skwarnicki:X3872PLep2003}.
But this mass was about 60~MeV above potential model predictions for a $1D$ charmonium state. The striking ``coincidence'' that, within the uncertainties, the state was at the $D\bar{D}^{0*}$ threshold was highly suggestive that this might be a $D^0\bar{D}^{0*}$ ``molecule''~\cite{Close:2003MolecHybrid,Braaten:2003X3872Molec,Tornqvist:2004X3872Mokc,Guo:2017HadMolec}. 

In contrast to the $\Theta$'s fate,
CDF internally confirmed Belle's peak eight days after their  arXiv posting.
CDF already had a pre-selected and reconstructed Run II $J/\psi$$\pi^+\pi^-$ sample  waiting on disk  to  continue an internal study of Run I data, where a weak signal around 3870 was seen in 1994.
Using 220 pb$^{-1}$ of $\bar{p}p$ collisions at $\sqrt{s}$$=$1.96~TeV CDF found
$730 \pm 90$ candidates (exceeding  10$\sigma$) and  measured the mass to be $3871.3\pm0.7 \pm0.4$~MeV. 
CDF presented its confirmation that September at the {\sc 2nd Workshop on Quarkonium}~\cite{CDF:3872Bauer2Quark}, and submitted a paper to PRL in December~\cite{CDF:3872-2003}.
Confirmation papers from D\O~\cite{D0:X3872-2004} and {\sc BaBar}~\cite{BaBar:X3872-2004} followed in May and June of 2004.
All experiments were in good accord.
This was an exciting discovery, but it was difficult to exclude a conventional explanation---maybe the charmonia mass predictions were off?

There was no clear consensus as to the nature of the $X(3872)$---evidence pointed in different directions:
\begin{itemize}
    \item {$\boldsymbol{m(D\bar{D}^{0*})-m(3872)}$}: 
    It was striking that the $X(3872)$ mass was, within uncertainties, consistent with the $D\bar{D}^{0*}$ mass.
    Currently  they  agree to better than 200~keV (within $1\sigma$)\cite{ParticleDataGroup:2024},
    pointing to a $D\bar{D}^{0*}$ molecule, or perhaps to an artifact of dynamic threshold effects.
%
%
    \item {$\boldsymbol{m(\pi^+\pi^-)}$}:
    Belle~\cite{Belle:X3872}, and later CDF~\cite{CDF:3872PiPi2005,Yi:2003st}, found evidence that the dipion mass spectrum in $J/\psi$$\pi^+\pi^-$ comes (predominantly) from $\rho$ mesons, whereas such isospin breaking is not natural for $c\bar{c}$ states.
    \item {\bf Hadronic production:} The $B$-factories found the $X(3872)$ in $B$ decays, but hadron colliders may directly produce the particle.
    An immediate suspicion was that large non-$B$ production at a hadron collider counted against an exotic interpretation~\cite{Chao:MysteryMeson}.
    In 2004 CDF found that its signal was dominantly directly produced:
    $16.1 \pm 4.9 \pm 2.0$\%  came from $B$'s~\cite{Bauer:X3872Life2004}.
    (Similarly, CMS found a fraction of  $26.3 \pm 2.3 \pm 1.6\%$ in $\sqrt{s}$= 7~TeV $pp$ collisions~\cite{CMS:3872xSec2013}.) 
    It was later quantified that CDF's direct $X(3872)$ production was orders of magnitude too large for a molecule~\cite{Bignamini:2009sk}.
    But it has since been argued that  the effects of $D^0\bar{D}^0$ rescattering of an $S$-wave threshold resonance were not accounted for, and this boosts the production by orders of magnitude, so it is claimed that a molecular interpretation  can accommodate hadronic production data~\cite{Artoisenet:2009wk,Braaten:2020iqw}.
\end{itemize}

There has been much progress  since these early days.
The $X(3872)$ is now seen in seven modes~\cite{ParticleDataGroup:2024}---and some  modes predicted by particular models  have not been seen.
It has also been observed in $p$-Pb~\cite{LHCb:pPb2024} and  Pb-Pb~\cite{CMS:3872HIon2019} collisions---providing a radically new arena to study its production.
Over the years its  $J^{PC}$ has been whittled down to $1^{++}$~\cite{LHCb:3872JPC2013,LHCb:3872JPC2015}---prompting its renaming to  $\chi_{c1}(3872)$~\cite{ParticleDataGroup:2024}.

LHCb found its natural width is extremely small: $0.96 ^{+ 0.19}_{-0.18} \pm 0.21$ ~MeV with a Breit-Wigner lineshape~\cite{LHCb:3872width2020b}, or $0.22 ^{+0.07+0.11}_{-0.06-0.13}$~MeV with the Flatt\'e formalism~\cite{LHCb:3872width2020a}.
The lineshapes were not distinguishable by the data, but the analytic structure of the Flatt\'e  amplitude favored a quasibound $D\bar{D}^{0*}$, with  a compact $c\bar{c}$ component $<\!33\%$~\cite{LHCb:3872width2020a,LHCb:3872width2020b}.
This has been contested~\cite{Esposito:3872Lineshap2021,Baru:3872Lineshap2021}, but the debate has moved away from either-or to what is the fraction of molecular vs compact.





\subsection{A theoretical interlude}

A fuller picture of the challenge in unravelling the nature of the $X(3872)$, and other heavy exotic candidates, may be appreciated by noting the profusion of exotic proposals.
These fall into a few classes:
\begin{itemize}
 \item Four-quark systems:
    \begin{itemize}
         \item Meson-Meson molecules~\cite{Dalitz:1959HadMolec,Voloshin:1976CharmMolec}: e.g. $X(3872)$ as loosely bound $D$--$\bar{D}^{0*}$ system~\cite{Close:2003MolecHybrid,Braaten:2003X3872Molec,Tornqvist:2004X3872Mokc,Guo:2017HadMolec};                        
         \item $c\bar{c}$ core $+$ molecular component~\cite{vanBeveren:2020eis};  
         \item Hadro-charmonium: compact $c\bar{c}$ embedded in an excited $q\bar{q}$ cloud~\cite{Dubynskiy:2008HadroCharm};  
         \item Tetraquarks: tightly bound $c\bar{c}q\bar{q}$ system, usually as a diquark pair $cq$-$\bar{c}\bar{q}$~\cite{Iwasaki:4c1975,Anselmino:1992Diquarks,Maiani:2004DiQuark};                                                 
    \end{itemize}  
  \item Hybrids: $c\bar{c}$ core $+$  valance gluon~\cite{Jaffe:Exotics1975,Giles:1976Hybrids,Horn:1977Hybrids,Meyer:2015eta,Oncala:2017Hybrid,Wan:2020fsk};   
  \item Threshold effects:            
 coupled-channel interactions~\cite{Swanson:2014CouplChann,Dong:2020nwy-Duplct}; 
 threshold cusps~\cite{Bugg:2008Cusp,Guo:2019twa};
  triangle singularities \cite{Szczepaniak:2015eza,Guo:2019twa}; and
 Pomeron exchange       \cite{Gong:2020bmg}.
\end{itemize}
The various four-quark options are, of course, idealizations of pure cases whereas there is really a  dynamical continuum  among them, and the wave function for an exotic is likely a mixture of
components.
A weakly bound molecular state, with some small-ish compact $c\bar{c}$ component, seems to be a heavy favorite for the $X(3872)$.

Exotic speculations were triggered by the fact that the $X(3872)$ mass significantly conflicted with potential model expectations~\cite{Belle:X3872}.
However, since those early days it has been better appreciated that these  charmonium mass predictions can be substantially shifted by interaction with open flavor thresholds, and
there have been continued efforts to interpret the $X(3872)$ as a conventional $c\bar{c}$ state~\cite{Suzuki:2005X3872cc,Meng:2007X3872cc,Ferretti:2013X3872cc}.
Thus it has been difficult to rigorously exclude conventional interpretations.

\begin{figure}[tb]
     \centering\resizebox{0.32\linewidth}{!}{\includegraphics{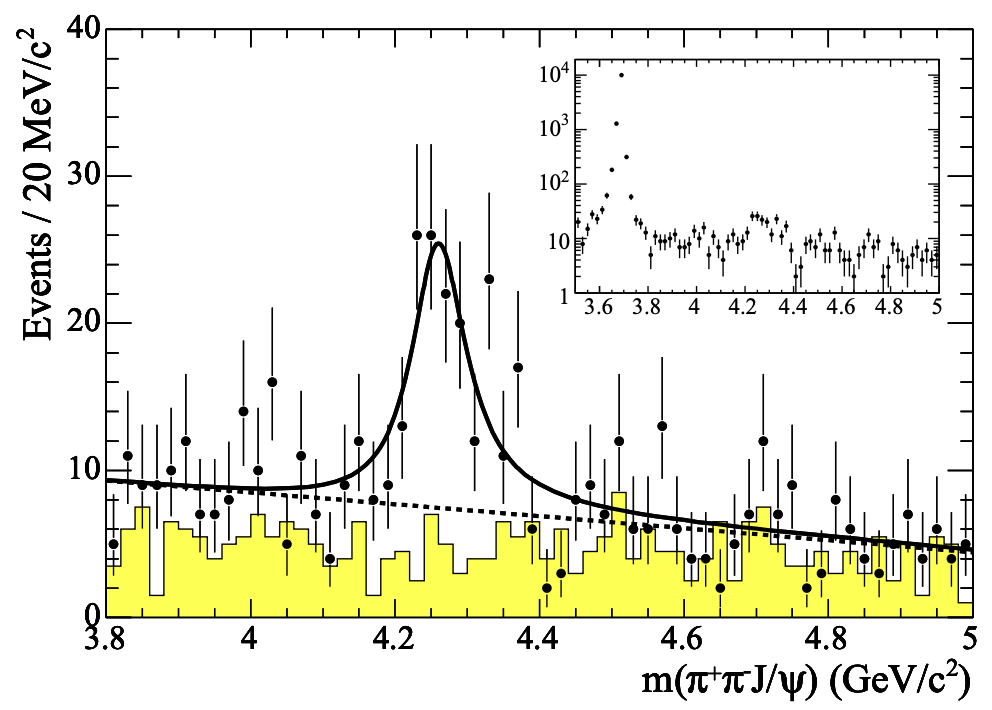}}
     \resizebox{0.32\linewidth}{!}{\includegraphics{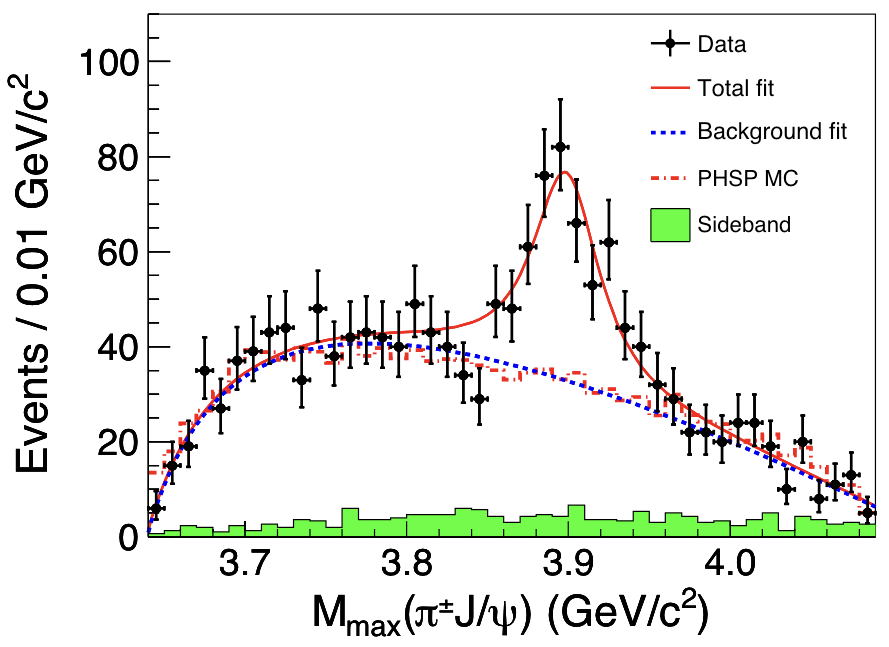}}
     \resizebox{0.32\linewidth}{!}{\includegraphics{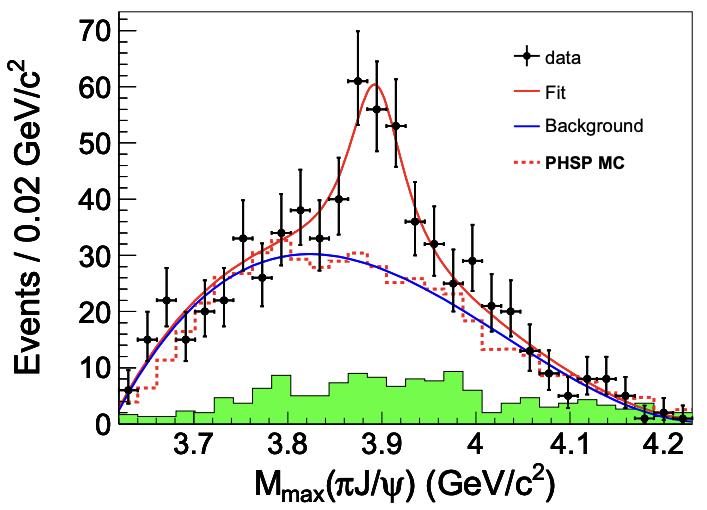}   } 
     \caption{The $J/\psi$$\pi^+\pi^-$ spectrum from \textsc{BaBar} showing the charmonium-like $Y(4260)$ (left)~\cite{BaBar:2005hhc}. 
     The $J/\psi \pi ^{\pm}$ spectrum from BESIII (middle) ~\cite{BESIII:2013ris}, and Belle  (right)~\cite{Belle:2013yex}, showing the $Z_{c}^{+}(3900)$ peaks. 
     }
     \label{fig:Z3900compare}
 \end{figure}

\subsection{Establishment of exotic hadrons: Discovery of charged charmonium-like states}
\indent

Discovery of the  \(X(3872)\) meson  opened the floodgates, and discovering exotic candidates seemed to become almost routine~\cite{exotic23}.
It was difficult to ``prove'' a state was exotic---unless the candidate's quantum numbers, such as its  $J^{PC}$, are forbidden for $q\bar{q}$ or $qqq$ systems.
Charmonium-like states have one gross property prohibited in the quark model: $c\bar{c}$ states must be neutral.
But, charged charmonium-like states are not easy to come by.

An early search by \textsc{BaBar}  in 2004 using 234 million $B\bar{B}$ decays came up empty handed~\cite{BaBar:X3872Charged2004}.
In 2008 Belle laid claim to the first charged charmonium-like state in  657 million $B\bar{B}$ events with a narrow peak ($6.5\sigma$) in the $\psi(2S)$$\pi^+$ mass distribution from 
$\bar{B}^0 \to\,$$\psi(2S)$$ K^- \pi^+$ and
$B^+ \to\,$$\psi(2S)$$K_S^0 \pi^+$ decays
at  $4433 \pm 4 \pm 2$~MeV, 
with a width of
$45 ^{+18+30}_{-13-13} $ MeV, 
named  $Z(4430)^+$ (Fig.~\ref{fig:Z4430})~\cite{Belle:2007hrb}. 
%
However, \textsc{BaBar} investigated whether mass and angular reflections of the $K\pi^+$ system could describe the structure of the $\psi(2S)$$\pi^+$ mass distribution.
While they could not exclude the possibility of a $Z(4430)^+$, they found no need for it~\cite{BaBar:NoZ2008}.
Belle responded with a 2D Dalitz-plot analysis confirming their original mass but a much larger width:
$4443 ^{+15+10}_{-12-13} $ and
$107  ^{+86+74}_{-43-56} $~MeV~\cite{Belle:4430Dalitz2009}. 
Then in 2013 Belle performed a more sensitive full 4D-amplitude analysis on 2k
$B^0 \to \psi(2S) K^+ \pi^-$ decays, which saw some further parameter shifts---$4485 \pm 22 ^{+28}_{-11}$ and
$ 200 ^{+41+26}_{-46-35} $~MeV---and $J^P = 1^+$  was favored~\cite{Belle:4430JPC2013}.  
The next year  LHCb confirmed Belle (also with 4D analysis, but with 25k $\psi(2S) K^+ \pi^-$ decays)  with more precise values:
$4475 \pm  7 ^{+15}_{-25}$ and
$ 172 \pm 13 ^{+37}_{-34}$~MeV, and determined its true resonant character with an Argand plot~\cite{LHCb:4430Resonant2014}. 

This fully settled the existence of the ``4430''  in  $\psi(2S) \pi^+$,  but before this state was resolved reports of  a \(Z_c(3930)^+\) state in $e^+e^- \to J/\psi \pi^+ \pi^-$ appeared from both BESIII and Belle experiments in 2013.
Years earlier
\textsc{BaBar} saw a charmonium-like exotic candidate $Y(4260)\to$$J/\psi$$\pi^+\pi^-$ (Fig.~\ref{fig:Z3900compare}, left) via initial state radiation (ISR), {i.e.}
$e^+ e^- {\to} \gamma_{ISR} J/\psi\pi^+\pi^-$~\cite{BaBar:2005hhc}, where its $J^{PC}$ must be $1^{--}$.  Inspired by the suspicion of its exotic nature and $J^{PC}$, BESIII searched $e^+e^-\to$$J/\psi$$\pi^+\pi^-$ data at an energy of 4.26~GeV---i.e.~at the Y(4260) mass.
In 2013 they reported a $J/\psi$$\pi^+$ signal of $307 \pm 48$ candidates ($8\sigma$) for a peak with a mass and width of 
$3899.0 \pm 3.6 \pm 4.9$ and
$46 \pm 10 \pm 20$~MeV (Fig.~\ref{fig:Z3900compare}, center)~\cite{BESIII:2013ris}.
Similarly Belle found a signal of $159 \pm 47 \pm 7$ candidates ($>5.2\sigma$) at a mass and width of $3894.5 \pm 6.6 \pm 4.5$ and $63 \pm 24 \pm 26$~MeV (Fig.~\ref{fig:Z3900compare}, right)~\cite{Belle:2013yex}.

Now one finally had conclusive evidence for non-$q\bar{q}$ mesons, with many other charged charmonium-like states to follow: $Z_c(3900)^+$, $Z_c(4020)^+$, $Z_c(4050)^+$\ldots;
 and even the bottomonium-like $Z_b(10610)^+$ and $Z_b(10650)^+$~\cite{ParticleDataGroup:2024}.

\subsection{Pentaquarks}
\indent

\begin{figure}[tb]
    \centering
    \resizebox{0.45\linewidth}{!}{\includegraphics{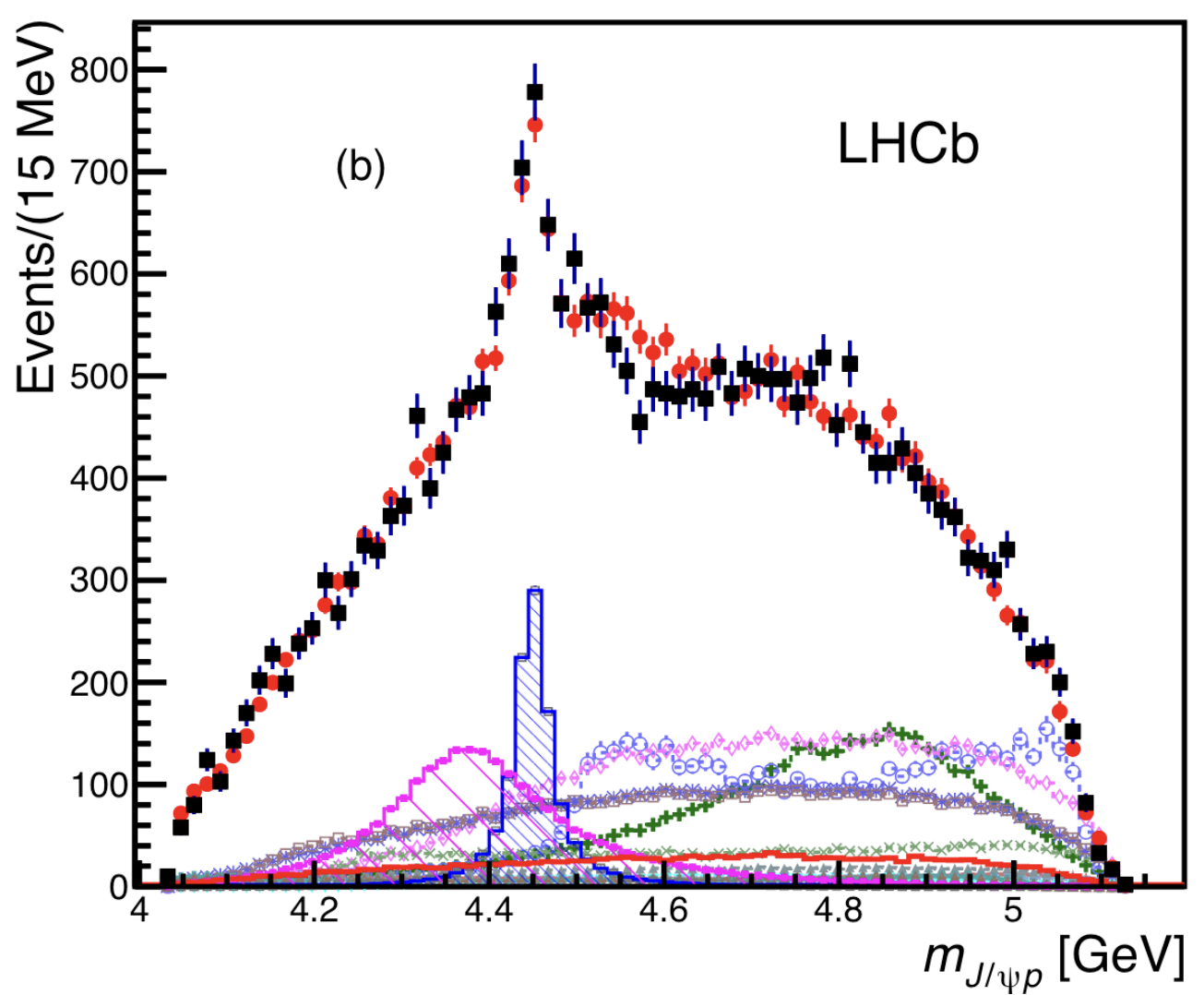}}
    \resizebox{0.4\linewidth}{!}{\includegraphics{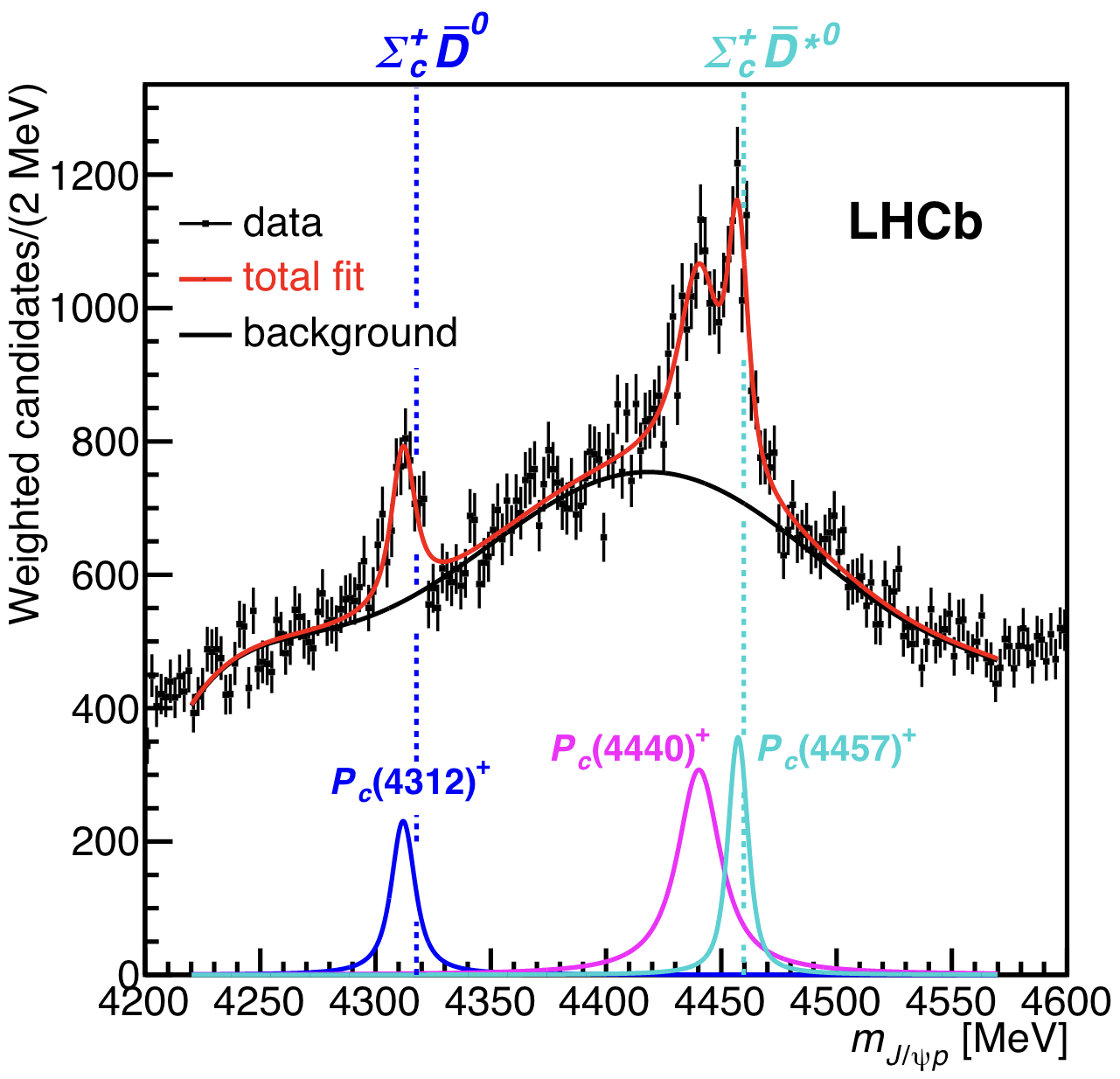}}
    \caption{Resonant structures in the $J/\psi p$ mass spectrum from LHCb. Two structures, named $P_c (4450)^+$ and $P_c (4380)^+$, were observed in a 2015 analysis with 3~fb$^{-1}$ (left)~\cite{LHCb:2015yax}. 
    A 2019 analysis with 6~fb$^{-1}$ (right) revealed that the $P_c (4450)^+$ was actually resolved into two overlapping narrow peaks, $P_c (4440)^+$ and $P_c (4457)^+$, 
    and a new narrow state, $P_c (4312)^+$, was also seen~\cite{LHCb:2019kea}.
     \label{fig:lhcbpentaquark}
   }
\end{figure}

Despite the $\Theta^+$ fiasco, the potential for pentaquarks was a sound idea, especially in light of the discovery of the $X(3872)$ and its companions.
 In 2015, LHCb used 3~fb$^{-1}$ of $\Lambda^{0}_{b} \to J/\psi K^{-} p$ decays, dominated by  $\Lambda^{*} \to K^{-} p$ resonances,  to search for  $J/\psi p$ structures. Two peaks were identified at $4380 \pm 8 \pm 29$ and $4449.8 \pm 1.7 \pm 2.5~\text{MeV}$, with respective widths of $205 \pm 18 \pm 86$ and  $39 \pm 5 \pm 19$~MeV (Fig.~\ref{fig:lhcbpentaquark}, left)~\cite{LHCb:2015yax}.
 The significance of each of these $P_{c}^{+}$ peaks was more than $9\sigma$.

With an increased signal efficiency for the  $\Lambda^0_b$,  LHCb obtained a sample of  $\Lambda^0_b \to J/\psi p K^-$  decays that was nine times larger than before. This larger sample revealed 
three hidden-charm pentaquark states: $P_c (4312)^+$, with a statistical significance of  $7.3\sigma$, and  the previous $P_c (4450)^+$ was resolved into two narrow overlapping peaks,  $P_c (4440)^+$ and $P_c (4457)^+$, with a significance of $5.4\sigma$ for the two-peak interpretation (Fig.~\ref{fig:lhcbpentaquark}, right)~\cite{LHCb:2019kea}.

However, LHCb could not confirm or refute the existence of the previously reported broad  $P_c(4380)^+$  state. This was because the newly observed peaks were so narrow that broad structures, such as the  $P_c(4380)^+$, were difficult to isolate in the new fit as they blended in with other contributions that vary more slowly across the  $m(J/\psi p)$  range. The fits with and without the broad  $P_c(4380)^+$  state both describe the data well, and so they could not discriminate between the two possibilities and the status of this broad structure remains open.

In  2021 LHCb reported the first evidence of a pentaquark  with a strange quark,  $P_{cs}(4459)^0$, in the  $J/\psi\Lambda^0$  channel from an amplitude analysis of the  $\Xi^-_b \to J/\psi\Lambda^0 K^-$ decay~\cite{LHCb:2020jpq}. This structure had a mass of  $4458.8 \pm 2.9 ^{+4.7}_{-1.1}~\text{MeV}$  and a width of  $17.3 \pm 6.5 ^{+8.0}_{-5.7} ~\text{MeV}$. Then, in 2022, LHCb announced another strange pentaquark, named $P_{c\bar{c}s}(4338)^0$, in an amplitude analysis of the $B^- \to J/\psi \Lambda^0 \bar{p}$ decay~\cite{LHCb:2022ogu}. The mass and width of this  candidate were measured to be $4338.2 \pm 0.7 \pm 0.4 ~\text{MeV}$ and $7.0 \pm 1.2 \pm 1.3 ~\text{MeV}$, respectively.

The states observed in the  $J/\psi p$  system, namely  $P_c(4312)^+$,  $P_c(4440)^+$, and  $P_c(4457)^+$, have minimal quark content of  $c\bar{c}uud$, consistent with hidden-charm pentaquarks. On the other hand, the  $P_{cs}(4459)^0$ and $P_{c\bar{c}s}(4338)^0$  observed in the  $J/\psi \Lambda$ have minimal quark content of  $c\bar{c}uds$, indicating the presence of strangeness in this pentaquark.
All candidates have, so far, only been reported by LHCb, and one looks forward to other experiments entering this arena.

\begin{figure}[tb]
    \centering
    \resizebox{0.55\linewidth}{!}{\includegraphics{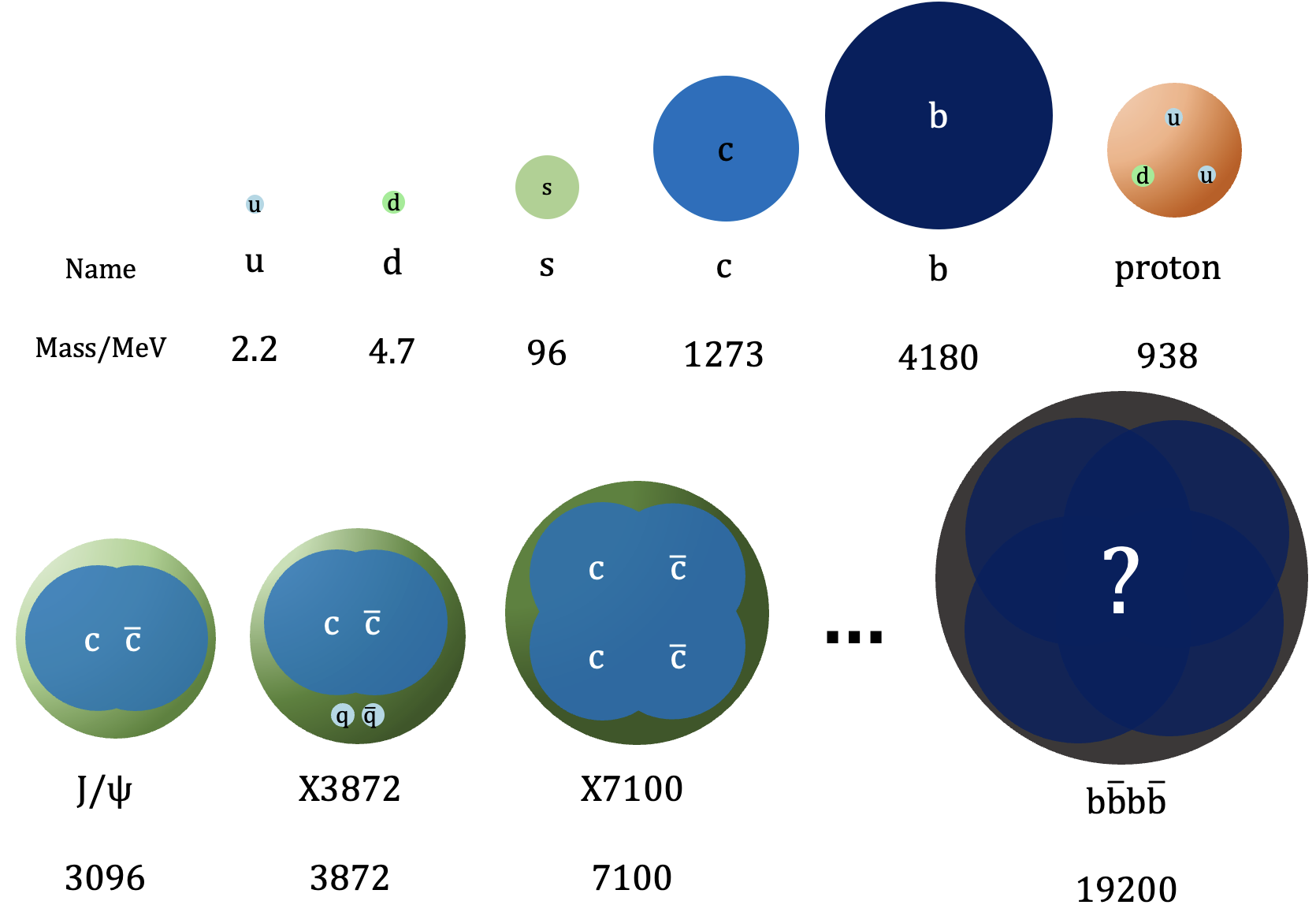}}
    \caption{Illustration of the scale of particle masses.
     \label{fig:MassCartoon}
   }
\end{figure}

\begin{table}[tb]
    \centering
    \caption{Summary of structures reported in the $J/\psi\phi$ mass spectrum. The significances are evaluated accounting for total (statistical-only) uncertainties.}
    \begin{tabular}{@{}lccccc@{}} 
    \hline
    Year   Experiment & Luminosity &\makecell[c]{Process and yield\\$B \rightarrow J/\psi\phi K$} & Mass (MeV) &  Width (MeV) &Signf.[$*\sigma$] \\
    \hline
    \hline
    2009  CDF\cite{CDF:2009jgo} &$2.7\, \mathrm{fb}^{-1}$ & $58\pm10$ & $4143.0\pm2.9\pm1.2$ & $11.7^{+8.3}_{-5.0}\pm3.7$  & $3.8$  \\
    
    2011  CDF\cite{CDF:4140-4140Obs2011,CDF:2011pep} &$6.0\,  \mathrm{fb}^{-1}$ & $115\pm12$ & \makecell[c]{ $4143.4^{+2.9}_{-3.0}\pm0.6$ \\ $4274.4^{+8.4}_{-6.7}\pm 1.9$ } & \makecell[c]{ $15.3^{+10.4}_{-6.1}\pm2.5$ \\ $32.3^{+21.9}_{-15.3}\pm 7.6$ }  & \makecell[c]{ $5.0$\\$3.1$ } \\ 
    
    2012  LHCb\cite{LHCb:2012wyi}& $0.37\, \mathrm{fb}^{-1}$ & $346\pm20$ & 4143.0 fixed& 15.3 fixed  \\
    
    2013  CMS\cite{CMS:2013jru}& $5.2\, \mathrm{fb}^{-1}$ & $2480\pm160$ & $4148.0\pm2.4\pm6.3$ & $28^{+15}_{-11}\pm19$  & $5.0$  \\
    
    2013  D0\cite{D0:2013jvp}& $10.4\, \mathrm{fb}^{-1}$ & $215\pm37$ & $4159.0\pm4.3\pm6.6$ & $19.9\pm12.6^{+1.0}_{-8.0}$  & $3.1$\\
    
    2014  \textsc{BaBar}\cite{BaBar:2014wwp}& $422.5\, \mathrm{fb}^{-1}$ & $189\pm14$ & 4143.4 fixed & 15.3 fixed  & $1.6$  \\
    
    2014 BESIII\cite{BESIII:2014fob}&  &$e^+e^{-} \to\gamma\phi J/\psi$ &\\
    
    2015  D0\cite{D0:2015nxw}& $10.4\, \mathrm{fb}^{-1}$ &$p\bar{p} \to J/\psi\phi + anything$ & $4152.5\pm1.7^{+6.2}_{-5.4}$ & $16.3\pm5.6\pm11.4$  & $4.7\, (5.7)$    \\
    \hline
    
    2016  LHCb\cite{LHCb:2016axx}& $3\, \mathrm{fb}^{-1}$&$4289 \pm151$&\makecell{$4146.5\pm4.5^{+4.6}_{-2.8}$\\$4273.3\pm8.3^{+17.2}_{- 3.6}$\\$4506\pm11 ^{+12}_{-15}$\\$4704\pm10^{+14}_{-2}$}&\makecell{$83\pm21^{+21}_{-14}$\\$56\pm11^{+8}_{-11}$\\$92\pm^{+21}_{-20}$\\$120\pm31^{+42}_{-33}$}&\makecell{8.4\\6.0\\6.1\\5.6}\\
    
    \hline
    2021  LHCb\cite{LHCb:2021uow}&$9\, \mathrm{fb}^{-1}$ &$24220 \pm 170$ &\makecell[c]{$4118 \pm 11 ^{+19}_{-36} $\\$4146\pm18\pm33$\\$4294 \pm 4 ^{+3}_{-6}$\\$4474 \pm 3 \pm 3 $\\$4626\pm16^{+18}_{-110}$\\$4684\pm7 ^{+13}_{-16} $\\$4694 \pm 4 ^{+16}_{-3}$} &\makecell{$162 \pm 21 ^{+24}_{-49}$\\$135\pm28^{+59}_{-30}$\\$53 \pm 5 \pm5$\\$77 \pm 6 ^{+10}_{-8}$\\$174\pm27^{+134}_{-73}$\\$126\pm15^{+36}_{-41}$\\$87 \pm 8 ^{+16}_{-6}$}&\makecell{13\,(16)\\4.8\,(8.7)\\18\,(18)\\20\,(20)\\5.5\,(5.7)\\15\,(15)\\17\,(18)}\\
    \hline
    \hline
    \end{tabular}
    \label{tab:jpsiphi-collection}
\end{table}

\begin{figure}[tb]
    \centering
\resizebox{0.44\linewidth}{!}{\includegraphics{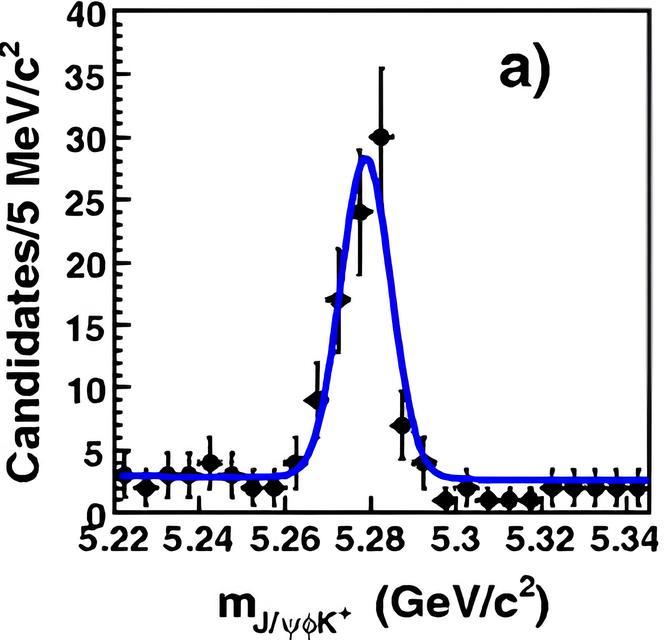}}
\resizebox{0.43\linewidth}{!}{\includegraphics{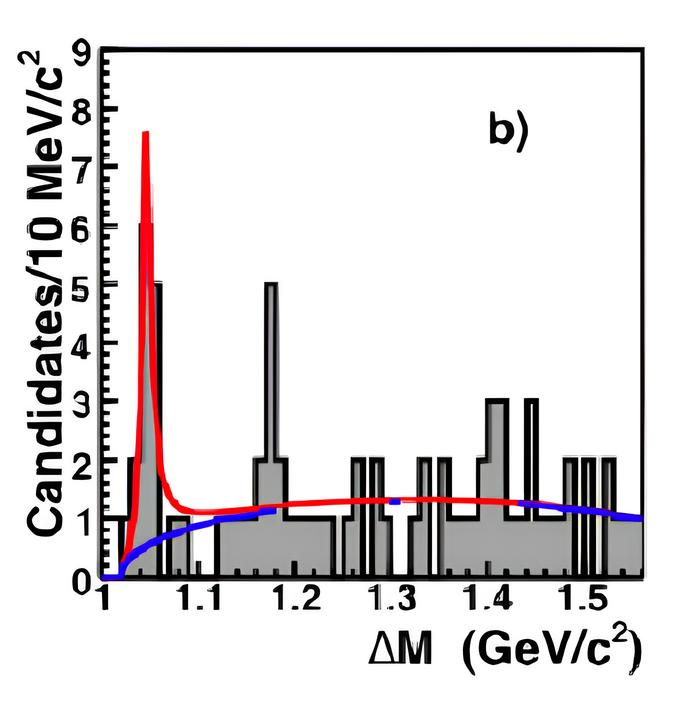}}
    \caption{The CDF reconstruction of $B^+ \rightarrow J/\psi \phi K^+$  ($\phi \rightarrow K^+ K^-$), showing a signal of $75 \pm 10$ $B$'s (left).
    The mass difference, $\Delta M$, between $\mu^{+}\mu^{-}K^{+}K^{-}$ and $\mu^{+}\mu^{-}$, for events in the $B^+$ mass window (right).
    The peak  ($14 \pm 5$ candidates) at $\Delta M$ of $1046.3 \pm 2.9$~MeV translates into a mass of $4143.0 \pm 2.9 \pm 1.2$~MeV~\cite{CDF:2009jgo}.}
    
    \label{fig:CDF4140}
\end{figure}

\section{The Road to All-Heavy Exotic Mesons---An Intermediate Step}
\indent

With unambiguous exotics in hand, and a flood of candidates, the theoretical challenge was how to understand their nature via their production, decay, and  properties. 
Compounding the problem, some argued that (at least some of) the exotic candidates may  be artifacts of dynamical effects of pair-production thresholds~\cite{Dong:2020nwy-Duplct,Guo:2019twa,Gong:2020bmg,
Wang:2020tpt}---many exotic candidates are close to such thresholds.
A basic difficulty in unraveling potential tetraquark structure is modeling systems containing light quarks.
Just as the heavy  charmonia spectrum brought clarity to the  quark model of mesons,  all-heavy tetraquarks hold similar promise.

Figure~\ref{fig:MassCartoon} shows a schematic comparison of the masses  of various quarks as sizes (except the top quark which is not spectroscopically relevant due to its  rapid decay), the proton, $X(3872)$, $X(7100)$, and a hypothetical $4b$ state.  
Hadrons with light quarks are  more complicated  to model due to their highly relativistic dynamics and non-perturbative binding. 
This is trivially illustrated by the fact  that a proton weighs 938~MeV, but its quarks collectively  weigh 
$\lesssim 10$~MeV, i.e. the quarks inside only account for 10\% of its mass.
This is due to the strong quark binding and their relativistic velocities.
In contrast, $\approx$70\% of the $X(7100)$ mass is accounted for by the four quarks inside (4*1.28=5.1~GeV).  
An all-heavy quark structure is much simpler  than a proton's.

\begin{figure}[tb]
    \centering
    \begin{minipage}[b]{0.51\textwidth}
\resizebox{0.99\linewidth}{!}{\includegraphics{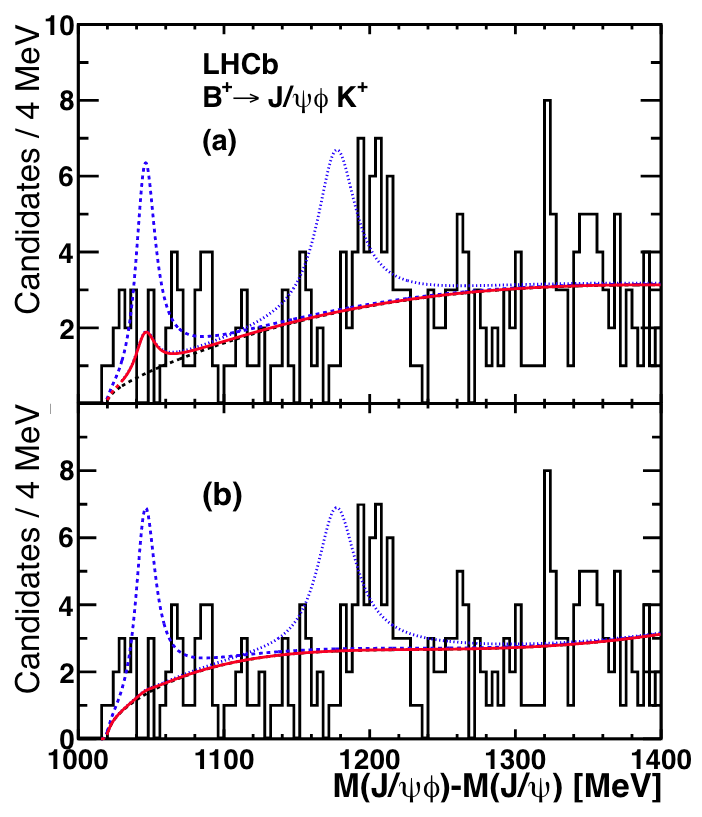}}
    \caption{The mass difference $M  (J/\psi \phi) - M (J/\psi)$ in $B^+ \to J/\psi \phi K^+$  decays from LHCb. The (a) and (b) fits are different background functions.~\cite{LHCb:2012wyi}.
    \label{fig:LHCb4140}
}
\end{minipage}
\hfill
\begin{minipage}[b]{0.46\textwidth}
    \centering
    \resizebox{0.94\linewidth}{!}{\includegraphics{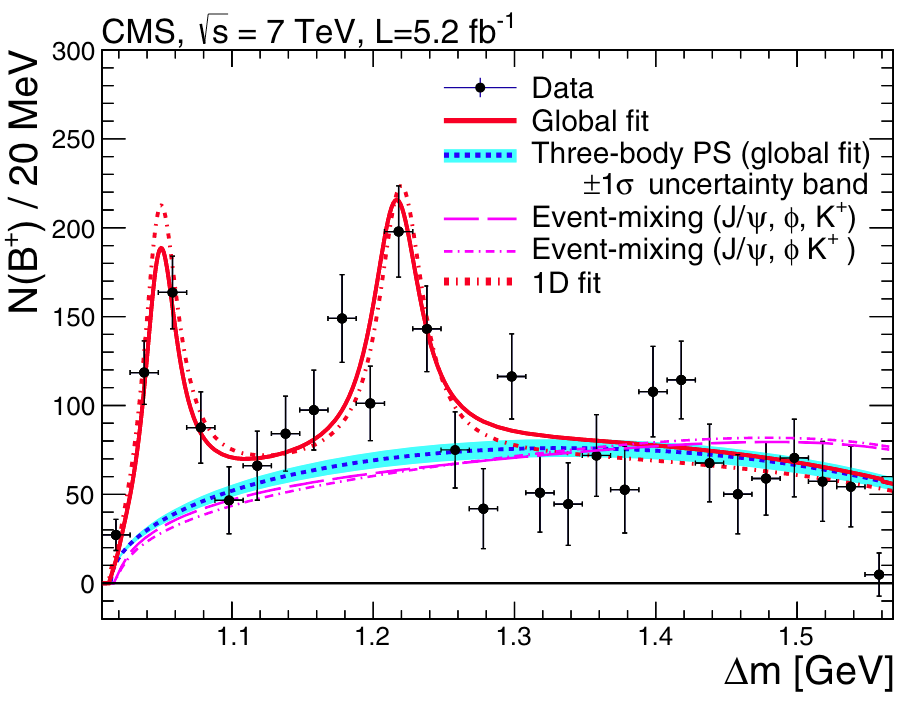}}\\
    \resizebox{0.94\linewidth}{!}{\includegraphics{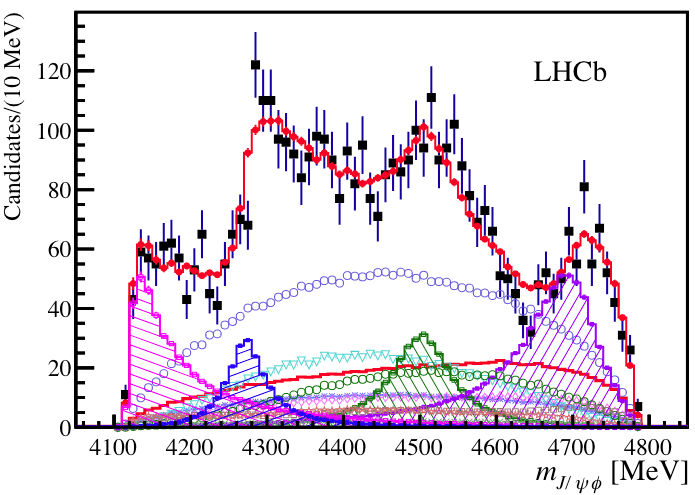}}
    \caption{The number of $J/\psi \phi$ candidates as a function of $\Delta m = m(\mu^{+}\mu^{-}K^{+}K^{-})- m(\mu^{+}\mu^{-})$ in CMS (top)~\cite{CMS:2013jru}.
    Distribution of $J/\psi \phi$ invariant mass for the $B^+ \to J/\psi \phi K^+$ data from LHCb  (bottom)~\cite{LHCb:2016axx}.}\label{fig:CMS4140LHCb}
\end{minipage}
\end{figure}

However,  hadrons with heavy quarks are more difficult to produce. 
The $s$ quark is  not  heavy, but its mass ($\sim$100~MeV) is about 40 times that of the $u$ and $d$ quarks, and so is relatively heavy. 
An intermediate step is to study
$c\bar{c} s\bar{s}$---a system with none of the lightest quarks---and hunt for the archetypical
$J/\psi$$(c\bar{c})  \phi (s\bar{s})$  channel.
Establishing  structure in $J/\psi\phi$ was somewhat convoluted---the inventory of candidates is tracked in Table~\ref{tab:jpsiphi-collection}.

In 2009 CDF made the first advance with  evidence of a $J/\psi \phi$ structure, labeled $Y(4140)$. 
 A signal of $14 \pm 5$ candidates ($3.8\sigma$ global signif.) was seen in 2.7~fb$^{-1}$ of $\bar{p}p$ data
(Fig.~\ref{fig:CDF4140}), for a mass 
of $4143.0 \pm 2.9 \pm 1.2$~MeV and width of $11.7 ^{+8.3}_{-5.0} \pm 3.7$~MeV\cite{CDF:2009jgo}.
The $Y(4140)$ is an analogue of the $X(3872)$ where a $q\bar{q}$ pair is replaced by the heavier $s\bar{s}$.
Belle found no evidence for the $Y(4140)$ in $B^+ \to J/\psi \phi K^+ $, but this did not contradict CDF~\cite{belle:2010}. 
Instead, they found evidence for $X(4350)$ in the $\gamma\gamma \to J/\psi \phi$ process~\cite{Belle:No41402009}.
%
In 2011, with 6~fb$^{-1}$, CDF confirmed the $Y(4140)$ with a significance of 5$\sigma$, and with parameters:
$4143.4 ^{+2.9}_{-3.0} \pm 0.6$ and $15.3 ^{+10.4}_{-6.1} \pm 2.5$~MeV~\cite{CDF:4140-4140Obs2011}.
A second structure, $Y(4274)$, was also identified with $22 \pm 8$ candidates 
(global significance $\approx 3.1\sigma$), with a mass and width of $4274.4 ^{+8.4}_{-6.7} \pm 1.9$ and $32.3 ^{+21.9}_{-15.3} \pm 7.6$~MeV.
However, in 2012, LHCb failed to confirm  these states, their mass distributions are shown in~Fig.~\ref{fig:LHCb4140}~\cite{LHCb:2012wyi}. 
These results diverged from CDF's results at the 2.4$\sigma$ level, and
publication of  CDF's paper was delayed until 2017~\cite{CDF:2011pep}.
The situation  at time was reviewed in Ref.~\cite{YI:2013iok}.

Resolution came in  2013,  when CMS  confirmed CDF's results using 5.2~fb$^{-1}$, and measured a  mass and width:
$4148.0 \pm 2.4 \pm 6.3$ and $28 ^{+15}_{-11} \pm 19$~MeV (Fig.~\ref{fig:CMS4140LHCb}, top)~\cite{CMS:2013jru}. 
Additionally, another structure was seen at a mass of $4313.8 \pm 5.3 \pm 7.3$ MeV, but its statistical significance could not be reliably determined. 
%
In the same year, D\O\,  offered further support for the $Y(4140)$ resonance ($3.1\sigma$) and indicated the presence of another one ($47 \pm 20$ candidates) at a mass of $4328.5 \pm  12.0$~MeV~\cite{D0:2013jvp}, consistent with a prior CMS result~\cite{CMS:2013jru}. 
In contrast to these successes, \textsc{BaBar} searched for $J/\psi \phi$ in 2014 but found nothing significant~\cite{BaBar:2014wwp}.
{
BESIII and D\O\, also explored new processes  for the $Y(4140)$. 
In 2014, BESIII searched $e^+e^-\to\gamma J/\psi\phi$ for the $Y(4140)$   at different center-of-mass energies, but no significant signal was observed\cite{BESIII:2014fob}. 
In 2015, D\O\,  searched for the inclusive production of the $Y(4140)$ in hadronic collisions and found strong evidence for its direct production\cite{D0:2015nxw}.
}

In 2016, LHCb performed an amplitude analysis of $B^+ \to J/\psi\phi K^+$, and established the $Y(4140)$ and $Y(4274)$~\cite{LHCb:2016axx}. However, the width of $Y(4140)$ was found to be substantially larger than before. The high $J/\psi\phi$ mass region was investigated for the first time with good sensitivity, revealing two significant  $0^{++}$ resonances: $X(4500)$ and $X(4700)$; and renamed all $J/\psi \phi$ 
 states as X states. 
In 2021, LHCb improved their amplitude analysis using a signal six times larger than before, providing further insights into the $Y(4140)$ and otherstructures\cite{LHCb:2021uow}. 
The study reaffirmed the  previous $X \to J/\psi\phi$ states  with higher significance and confirmed quantum number assignments. 
An additional $1^+$ state, $X(4685)$, was identified with a relatively narrow width around 125~MeV and high significance. 
Additionally, a novel state  $X(4630)$ was unveiled with a significance of 5.5$\sigma$. This state favors a spin-1 assignment over 2 at  3$\sigma$,  rejecting other $J^P$  at 5$\sigma$.  
LHCb's pioneering study  also made the first observation of states decaying to $J/\psi K^+$---charged states with the new quark content $c\bar{c}u\bar{s}$.
We now have a large bounty of charmonium-like states with strange quarks for model builders to study.

\section{Structures in the Di-Charmonium Channel}
\indent

The dream of all-heavy tetraquarks may finally have turned into reality in 2020 with LHCb's  observation of  $X(6900){\,\to\,}$$J/\psi\,J/\psi$ decays~\cite{LHCb:2020bwg,ATLAS:2023bft,CMS:2023owd}.
It may be surprising that consideration of such a rare beast as an all-charm tetraquark dates all the way back to Yoichi Iwasahi in 1975~\cite{Iwasaki:4c1975}, just months after the $J/\psi$ discovery!
This was founded on a misguided interpretation that the excited $\psi$ states might be $c\bar{c}q\bar{q}$ tetraquarks, which logically also lead to predicting an all-charm tetraquark around 6.2~GeV.
Iwasahi seemed (briefly) vindicated by interpreting 
the E288 (FNAL) experiment's 
newly discovered ``Upsilon'' (a.k.a ``Oops-Leon'') particle around 6~GeV~\cite{Hom:OpsLeon1976} 
as a four-charm state~\cite{Iwasaki:4c1976}.
Alas, this ``Upsilon'' fizzled~\cite{Yoh:Upsilon1997}.
Another early stab at $4c$ states was by Kuang-Ta Chao who, in 1981, speculated that apparent structure in the elevated cross sections seen in the $e^+e^-$ continuum [the $R(e^+e^-\to{\rm hadrons})$ ratio] over the 6-7~GeV range~\cite{Barnett:eeR1980} might be due to $P$-wave $(cc)$-$(\bar{c}\bar{c})$ states~\cite{Chao:4c1980}.

\begin{figure}[tb]
    \centering
    \resizebox{0.5\linewidth}{!}{\includegraphics{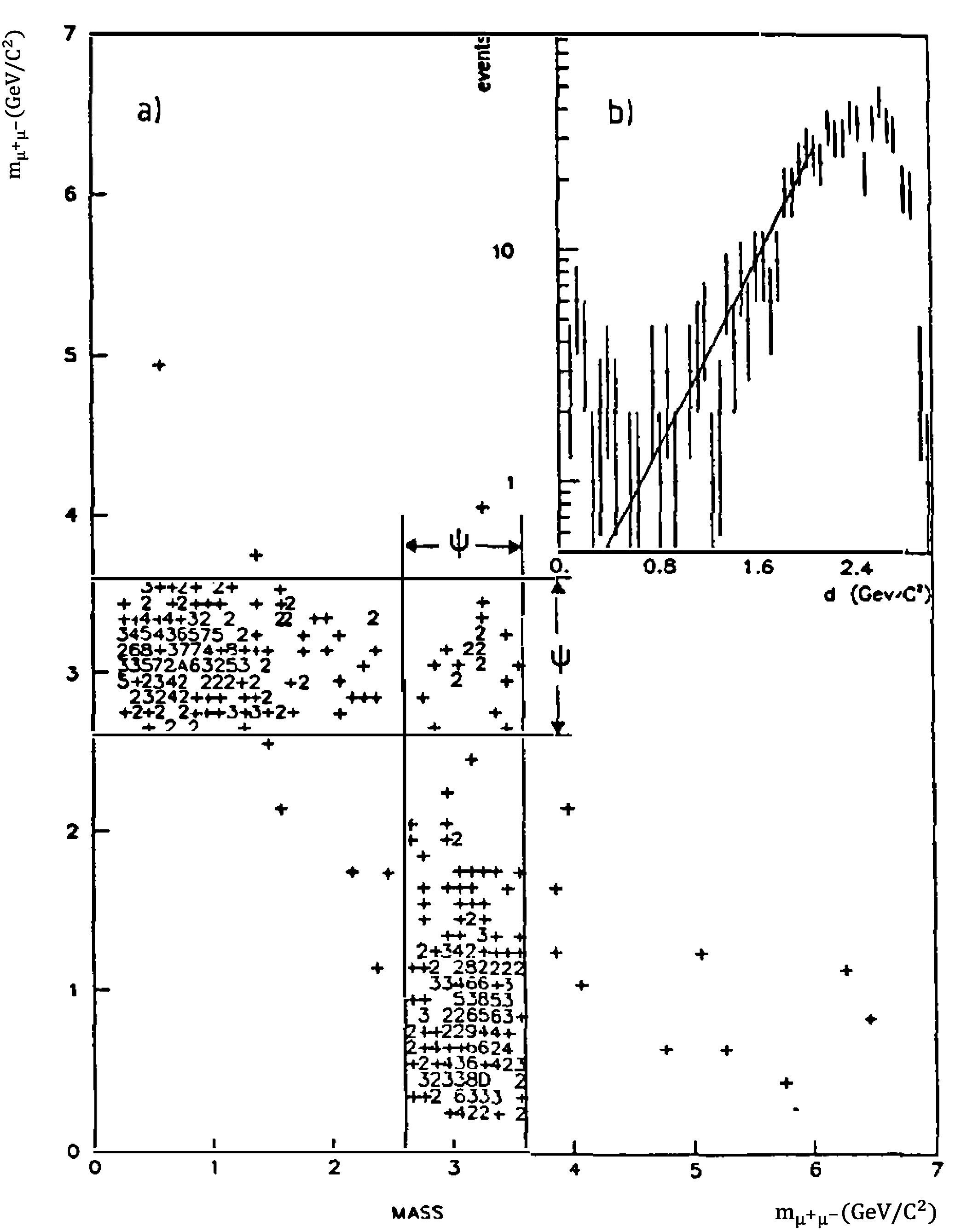}}
    \caption{Distribution of the two dimuon mass combinations in $\mu^+ \mu ^-\mu^+ \mu ^- $ events from NA3 in 1985\cite{NA3:JJprod1985}}
    \label{fig:NA3}
\end{figure}

The $J/\psi\,J/\psi$ channel is a prime discovery route to four-charm states.
Exploration of the  di-$J/\psi$ channel was pioneered by the NA3 experiment at CERN in 1982~\cite{NA3:JJprod1982,NA3:JJprod1985}, where they reported 13 $J/\psi\,J/\psi$ events (Fig.~\ref{fig:NA3}).
A new era opened with the advent of high luminosity hadron colliders where the relentless growth of $J/\psi\,J/\psi$ samples---116 pairs for LHCb in  
2012~\cite{LHCb:JJProd2011}, 
    1k for CMS  in 2014~\cite{CMS:JJprod2014}
    and LHCb    in 2017~\cite{LHCb:JJProd2016}---enabled  studies of $J/\psi\,J/\psi$ production.
CMS and LHCb had also showed crude $J/\psi\,J/\psi$ mass distributions, but the samples were  still small and the binning of the spectra were too large  ($>1$~GeV)  to be spectroscopically interesting.

Finally, in 2020,  with thirty times the data, LHCb  found the \(X(6900)\) structure~\cite{LHCb:2020bwg}---providing the first candidate for an all-heavy exotic.
It was confirmed by ATLAS~\cite{ATLAS:2023bft} and CMS~\cite{CMS:2023owd}.
Despite this consensus, the di-$J/\psi$ spectrum presents a  more interesting and complex situation: with two additional structures claimed by CMS, $X(6600)$ and $X(7100)$~\cite{CMS:2023owd}, and an unexplained  threshold feature. This spectrum promises more lessons.

\subsection{The $J/\psi\,J/\psi$\ mass spectrum}
\indent

%

\begin{figure}[tb]
    \centering
    \resizebox{0.27\linewidth}{!}{\includegraphics{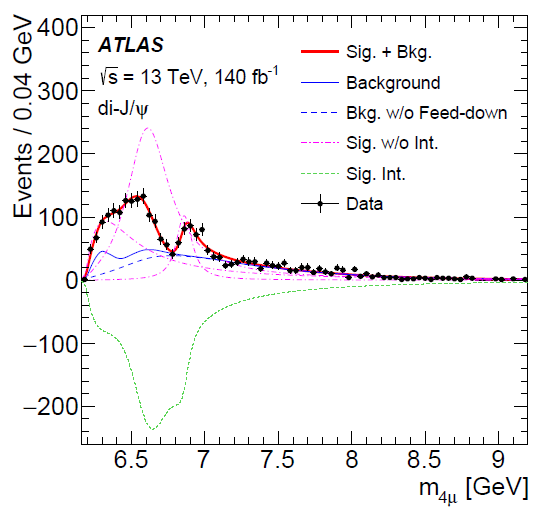}}
    \resizebox{0.33\linewidth}{!}{\includegraphics{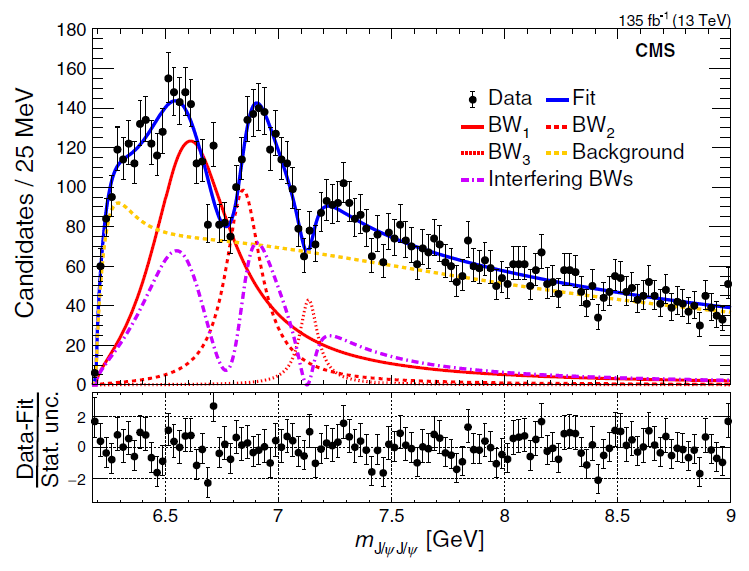}}
    \resizebox{0.38\linewidth}{!}{\includegraphics{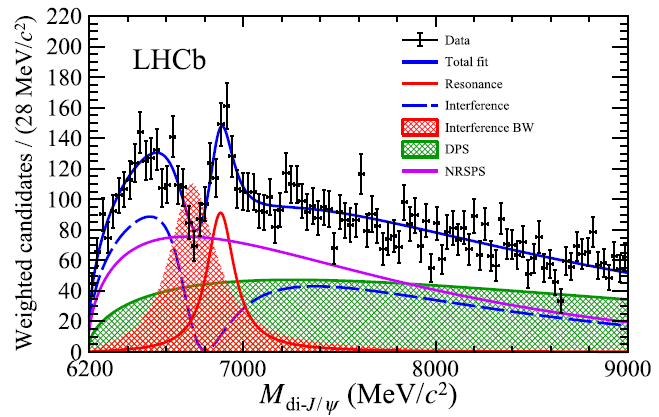}}

    \caption{The  $J/\psi\,J/\psi$ mass spectra from ATLAS with Model~A fit (left)~\cite{ATLAS:2023bft}, CMS interference fit (center)~\cite{CMS:2023owd}, and LHCb's  model II fit (right)~\cite{LHCb:2020bwg}.
    These are the experiments' interference fits, which their data prefer.
    \label{fig:JJspectra}
    }
\end{figure}

The  $J/\psi\,J/\psi$ samples used were:
140~fb$^{-1}$ (2015-2018) at $\sqrt{s}$=13~TeV for ATLAS,
135~fb$^{-1}$ (2016-2018) at $\sqrt{s}$=13~TeV for CMS, and 
9~fb$^{-1}$ 
(2011-2018) 
at  $\sqrt{s}$=7, 8, and 13~TeV for LHCb.
The  integrated luminosity for LHCb is smaller than the others but it still has similar spectroscopic reach as the other two because LHCb is designed for small-angle forward physics where $p_T$ cuts are much lower, and the production cross sections are larger.
Figure~\ref{fig:JJspectra} shows the $J/\psi\,J/\psi$ spectra from the three experiments.
The same broad features are seen---with structure, of {\it some} sort, immediately apparent.
The data rapidly rise at threshold, forming a broad ``hump'' peaking around 6500~MeV, then dropping into a pronounced ``dip'' around 6750~MeV before rising to a relatively narrow peak around 6900~MeV, and then tapering off at high masses.
There is also a small sign of a possible structure around 7200~MeV in the CMS and LHCb spectra.

The main background processes  are (non-resonant) single-parton scattering (NRSPS), and double-parton scattering (DPS)~\cite{Sjostrand:DoublePart1987,CDF:1993DoublePart, CDF:1997DoublePart}.
The NRSPS process is the conventional single parton-parton interaction in a single proton-proton collision.
At the very high LHC luminosities, the normally rare process of two parton interactions in a single proton-proton collision can become a significant part of triggers,  like $J/\psi\,J/\psi$ production.
Within the fit ranges,
DPS dominates at high $J/\psi\,J/\psi$ masses, and NRSPS at low masses.
There is also a small ``combinatoric'' background where at least one $J/\psi$ was ``fake.''
These backgrounds are common to all three experiments, but are  modeled in somewhat different ways based on a combination of Monte Carlo and data.

Constructing fit models is usually straightforward and uncontroversial.
This was not the case for the $J/\psi\,J/\psi$ spectra, as each experiment offered two different principal models,  resulting in three major types of
incongruent  fit models, and thereby offering a variety of different potential masses for roughly the same features.
%
%
We use LHCb's fitting models as the starting point for our discourse, as they pioneered  modeling the $J/\psi\,J/\psi$ spectrum.

\subsubsection{Simple multi-resonance fits}
\indent

\begin{figure}[tb]
    \centering
    \resizebox{0.445\linewidth}{!}{\includegraphics{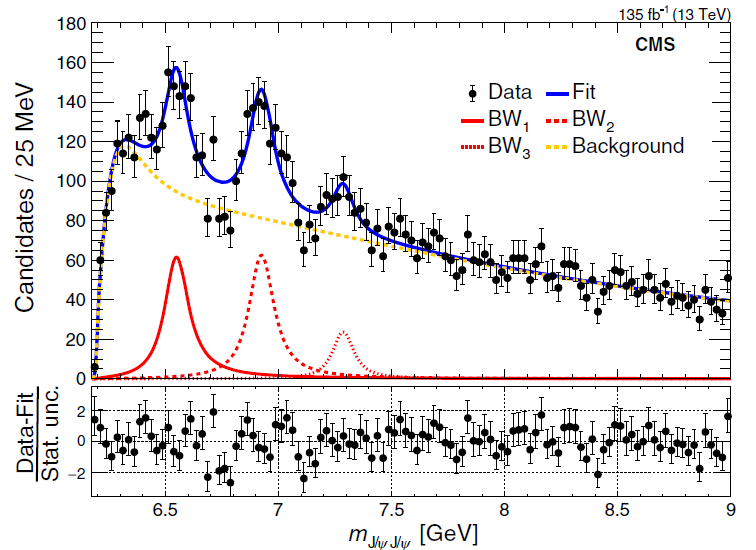}}
 \hfill
    \resizebox{0.51\linewidth}{!}{\includegraphics{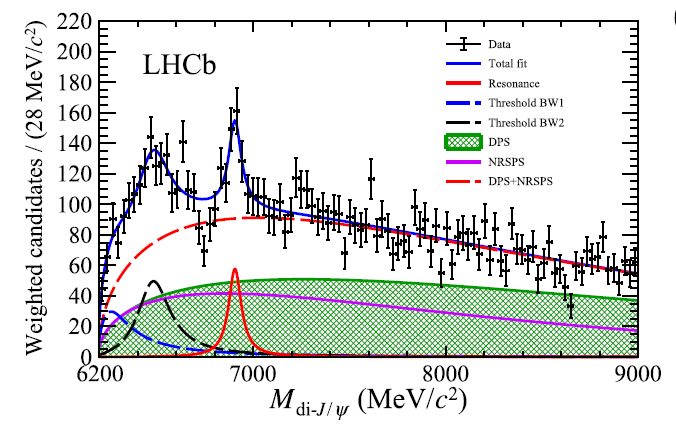}}  
    \caption{The  
    CMS~\cite{CMS:2023owd} (left) and LHCb~\cite{LHCb:2020bwg} (right) 
     $J/\psi\,J/\psi$ spectra fit without interference ({i.e.}, their respective ``no-interf.'' model and Model~II).
    \label{fig:JJspectra-NoInterf}
    }
\end{figure}

The Breit-Wigner (BW) lineshape was the function of choice to model the data---for physical resonances  and as an {\it ad hoc} shape for features of indeterminant origin.
It is clear (Fig.~\ref{fig:JJspectra})  that component(s) are needed to account for the broad ``hump'' around 6500~MeV, and for the 6900~MeV peak.
LHCb excluded the null hypothesis by $6.0\sigma$.
They tried to describe each of these structures with a single BW (along with the backgrounds), but found a fit $\chi^2$  probability of 1.2\% ~\cite{LHCb-JJ-Sup}.
The fit poorly described the threshold turn-on and high mass tail of the broad hump.
LHCb also described the hump with two BW's:
this three resonance fit, Model~I (Fig.~\ref{fig:JJspectra-NoInterf}, right; Table~\ref{tab:jpsijpsistructures})~\cite{LHCb:2020bwg}, was one of their two principal models.
%
The relatively narrow peak was fit with a mass and width of
$6905 \pm 11 \pm 7$ and $ 80 \pm 19 \pm 33$~MeV,
from $252 \pm 63$ candidates.
The significance of this signal exceeds $5\sigma$.
The fitted values of the two lower BW's were not reported, but their masses are around  6250  and 6500~MeV by visual inspection.
The threshold turn-on is described well, but the ``dip''  around 6750~MeV was not.  %
The fit probability was only 4.6\%.
LHCb also performed a supplementary four-BW fit with a BW for a small possible excess of events around 7200~MeV, but no parameters were reported as its significance was too low~\cite{LHCb-JJ-Sup}.

ATLAS also performed a three-BW fit, but they did not publish a plot or list resonance parameters~\cite{ATLAS:2023bft}.
However they stated this model was excluded  at a confidence level exceeding 95\%.
Given the similarity of spectra seen in  Fig.~\ref{fig:JJspectra}, we presume the fit's low probability is mainly due to the 6750~MeV dip.

CMS reported a four-BW fit as one of their main results:
two BW's described the threshold region, and one each for the peaks around 6900 and 7200~MeV---CMS included a BW  for any structure with a local significance above  $3\sigma$.
This fit is listed in Table~\ref{tab:jpsijpsistructures} as ``No-Interf'' (Fig.~\ref{fig:JJspectra-NoInterf}, left).
The  $X(6900)$  signal had $492 ^{+78} _{-73}$ candidates, and  parameters  consistent with LHCb's.
The CMS threshold turns on very fast, and evidently forms a brief shoulder before rising to a peaking structure around 6600~MeV, presenting a striking pattern of a total of three peaks, which CMS  called $X(6600)$, $X(6900)$, and $X(7100)$.
LHCb's spectrum (Fig.~\ref{fig:JJspectra-NoInterf}, right) does not appear inconsistent with this, but the effect is washed-out by their slower turn-on making their spectrum more amenable to alternate interpretations.
CMS's fit also had a  dip discrepancy around 6750~MeV like LHCb,
but there is another smaller dip discrepancy around 7150~MeV.
The parameters for the threshold BW were not reported because, as discussed later, CMS included it in the background.
The fit $\chi^2$ only for masses below 7.8~GeV yields a 9\% probability.
This fit cannot be dismissed statistically, but the 6750~MeV dip is the principal discrepancy.

\begin{table}[tb]
    \centering
    \newcolumntype{Y}{>{\raggedleft\arraybackslash}X}
    \caption{The principal results of fitting the $J/\psi\,J/\psi$ mass spectra by ATLAS, CMS, and LHCb.
    The experiments offer disparate models, and the parameters in one model cannot necessarily be compared to those in another. 
    To organize the results we take a phenomenological approach and categorize reported structures which are conflicted according to the range of their mass as ``Threshold'' or ``Intermediate'' BW's, along with the unambiguous ``$X(6900)$'' and ``$X(7100)$.''
%
    Masses and widths are in MeV.
      \label{tab:AllJJ}
    }
    \resizebox{\textwidth}{!}{
    \begin{tabular}{@{}cc|cc|cc|cc@{}} 
    \hline
    &   &  \multicolumn{2}{c|}{ATLAS~\cite{ATLAS:2023bft}} &  \multicolumn{2}{c|}{CMS~\cite{CMS:2023owd}} &  \multicolumn{2}{c}{LHCb~\cite{LHCb:2020bwg} }  \\
    &  & Model~A& Model~B & No-Interf. & Interf. & Model I & Model II \\
    \hline
Thresh. &  $m$    &  $6410 \pm 80 ^{+80}_{-30}$ 
         & ---
         & Unreported
         & Unreported
         & Unreported
         & ---
       \\
&$\Gamma$&  $590 \pm 350 ^{+120}_{-200}$ 
         & ---
         & Unreported
         & Unreported
         & Unreported
         & ---
        \\
    \hline        
Interm. &  $m$    &  $6630 \pm 50 ^{+80}_{-10}$ 
         & $6650 \pm 20 ^{+30}_{-20}$
         & $6652 \pm 10 \pm 12$
         & $6638 ^{+43+16}_{-38-31} $
         & Unreported
         & $6741 \pm 6$ \cite{LHCb-JJ-Sup}
       \\
&$\Gamma$&  $350 \pm 110 ^{+110}_{-40}$ 
         &  $440 \pm  50 ^{+ 60}_{-50}$ 
         &  $124 ^{+ 32}_{-26}  \pm 33$
         &  $449 ^{+230+110}_{-200-240}  $ 
         &  Unreported
         & $288 \pm 16$ \cite{LHCb-JJ-Sup}
        \\
    \hline        
$X(6900)$&  $m$    &  $6860 \pm 30 ^{+10}_{-20}$ 
         & $6910 \pm 10 \pm 10$
         & $6927 \pm 9 \pm 4$
         & $6847 ^{+44+48}_{-28-20} $
         & $6905 \pm 11 \pm 7$
         & $6886 \pm 11 \pm 11$
       \\
&$\Gamma$&  $110 \pm 50 ^{+20}_{-10}$ 
         &  $150 \pm 30 \pm 10$ 
         &  $122 ^{+ 24}_{-21}  \pm 18$
         &  $191 ^{+ 66+25}_{- 49-17}  $ 
         &  $ 80 \pm 19 \pm 33$ 
         &  $168 \pm 33 \pm 69$ 
        \\
    \hline        
$X(7100)$&  $m$    &  --- 
         & ---
         & $7287 ^{+20}_{-18} \pm 5$
         & $7134 ^{+48+41}_{-25-15} $
         & ---
         & ---
       \\
&$\Gamma$&  ---
         &  ---
         &  $ 95 ^{+59}_{-40} \pm 19$ 
         &  $97 ^{+ 40+29}_{- 29-26}  $ 
         & ---
         & ---
        \\ 
%
    \hline
    \end{tabular}
    \label{tab:jpsijpsistructures}      
    }
\end{table}

In summary, using multi-BW models all experiments found  similar fit behavior.
Notably the description of the threshold hump seems to require two BW's, and the dip around 6750~MeV was  poorly described across experiments.
The depth of the difficulty in describing the dip may be appreciated by noting that the dip in the data descends to about the same level as the fitted background (Fig.~\ref{fig:JJspectra-NoInterf}): {i.e.} there is little room for  additive contributions  from neighboring BW structures. 
One potential way to ameliorate this problem is to use multiple narrow BW's 
to model individual peaking structures, so that the BW's  could fall more steeply into the dip.
CMS alluded to this possibility~\cite{CMS:2023owd}, but since no such results were reported, we  pursue it no further.

\subsubsection{Interference with background fits}
\indent

To produce a dip in the fit to the level of the background, LHCb invoked  quantum mechanical interference.
A minimal way to employ interference is to have a {\it single} BW interfering with the NRSPS background.
It is unusual, but possible, for some NRSPS processes to have the same quantum numbers as, and maintain coherence with, a resonant structure.
LHCb performed such a fit, but it only had a $\chi^2$ probability of 2.8\% ~\cite{LHCb-JJ-Sup}.
Destructive interference  successfully generated a dip around 6750~MeV, but the threshold hump and the large and narrow 6900~MeV peak were poorly described.
More components are necessary to model their data.

LHCb's Model~II had BW's  for the 6900~MeV peak and an auxiliary BW  which interfered with the  NRSPS.
The results are shown in Fig.~\ref{fig:JJspectra} (right), and  in Table~\ref{tab:AllJJ}~\cite{LHCb:2020bwg}.
The auxiliary BW  fits to a mass of  $6741 \pm 6$~MeV---the location of the dip. The interference is maximal at the BW mass, and it contributes to the threshold hump by constructive interference below the peak.
This gives a good description of the dip, with a fit probability of 15.5\%.
The  $X(6900)$ does not participate in the interference, but since the ``background'' underneath it shifts, 
the resonance mass is lowered to $6886 \pm 11 \pm 11$~MeV, and the yield is increased to $784 \pm 148$ candidates.

ATLAS replicated LHCb's Model~II as their Model~B~\cite{ATLAS:2023bft}.
The results are given in Table~\ref{tab:AllJJ}.
The fit describes their data well, but
the  auxiliary BW parameters are not  very consistent with LHCb's: the ATLAS mass is almost 100~MeV lower, and the width is  $\sim$50\% larger.
Tension among  $X(6900)$ parameters is more modest.

CMS did not adopt Model~II, but for comparisons they did fit their data with it~\cite{CMS:2023owd,CMS-JJ-Sup,CMS:2022pas}.
%
The Model~II fit described CMS's 6750~MeV dip, and resulted in the masses and widths:  
$6736 \pm 38$ and $439 \pm 6$, and
$6918 \pm 10$ and $187 \pm 40$~MeV.
CMS's auxiliary mass matches LHCb's Model~II, but CMS's width was much broader (but it agrees with ATLAS's Model~B).
%
CMS's $X(6900)$ mass   is higher than LHCb's, but agrees with ATLAS's Model~B.
This situation appears to arise because the threshold turn-on of CMS, and to a lesser extant ATLAS, is  faster than for LHCb (compare in Fig.~\ref{fig:JJspectra}), and without a threshold BW the CMS fit must pull up the NRSPS to describe the turn-on (which is still poorly described for CMS up to 6.5~GeV: see pulls plot in Fig.~2 [right] of Ref.~\cite{CMS:2022pas}); but then to compensate for the large NRSPS the auxiliary BW must be increased and broadened (e.g. compare the backgrounds in Fig.~2 [right] of Ref.~\cite{CMS:2022pas} to Fig.~\ref{fig:JJspectra} [left]).
That BW parameters are sensitive to how   detectors sculpt the data indicates that the model
is probably not a good physical representation.


\subsubsection{Interfering resonance fits \label{Sec:InterfRes}}
\indent

Interference is able to describe the 6750~MeV dip, but
interference with NRSPS is not the only option.
Arguably it is more natural to arrange interference between the well-defined quantum states of resonances.

ATLAS's Model~A had interference among three resonances, two for the threshold hump and one for the $X(6900)$.
The results are shown in Fig.~\ref{fig:JJspectra} (left), and listed in Table~\ref{tab:AllJJ} (each peak $>5\sigma$).
The data is described well, and the $X(6900)$ parameters are consistent with LHCb's Model~II (NRSPS interference), and CMS (below).

CMS also adopted a multi-resonance interference fit  that mirrors the structure of their ``no-interf.'' model: a non-interfering ``background'' threshold BW plus three interfering BW's.
The fit result is shown in Fig.~\ref{fig:JJspectra} (center), and Table~\ref{tab:AllJJ} under ``Interf."
CMS advanced three physical structures,
\(X(6600)\), \(X(6900)\), and \(X(7100)\),
with local significances of 7.9$\sigma$, 9.8$\sigma$, and $4.7\sigma$ respectively.
The $X(6900)$ parameters are consistent with ATLAS's Model~A and LHCb's Model~II, but the latter has some tension with the CMS width. 
The 6750~MeV dip is well described, but the fit also favors an interference dip around 7100~MeV.

\begin{figure}
    \centering
    \resizebox{0.95\linewidth}{!}{\includegraphics{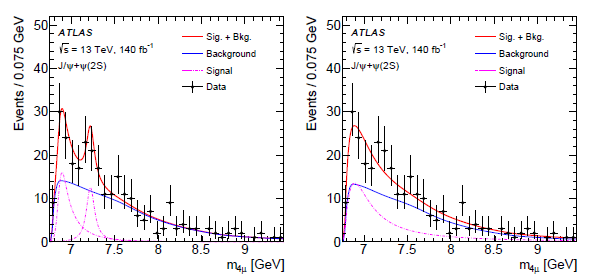}}
    \caption{The  $J/\psi  \psi(2S)$ mass spectra from ATLAS with Model~$\alpha$  (left) and Model~$\beta$ (right) fits~\cite{ATLAS:2023bft}. In addition to background, Model~$\alpha$ consists of two signal BW's, one fixed to the parameters of the $X(6900)$ from their $J/\psi\,J/\psi$ data and the second floats freely; Model~$\beta$ has a single floating BW.}
    \label{fig:ATLAS-J2s}
\end{figure}

\subsection{The $J/\psi  \psi(2S)$ spectrum}
\indent

It is natural to suspect that sufficiently massive $J/\psi\,J/\psi$ structures   will also  appear in the $J/\psi  \psi(2S)$ channel.
Estimates of the $X(6900)\to$$J/\psi  \psi(2S)$ branching fraction relative  to  $J/\psi\,J/\psi$ vary from 
${\sim}10$--110\%~\cite{PhysRevD.106.094019,Lu:DiQuarkSize2020}.

ATLAS searched the $\mu^+ \mu^- \mu^+ \mu^-$ channel and reported a substantial threshold excess, but again offered two alternate fit models~\cite{ATLAS:2023bft}.
In Model~$\alpha$ a threshold BW is fixed to the \(X(6900)\) parameters from their $J/\psi\,J/\psi$ data (Model~A), which is added to the background (plus the tails of the two sub-threshold BW's fixed to Model~A parameters),
and finally, and additional BW is included to describe a possible peak around 7200~MeV.
The result is shown in Fig.~\ref{fig:ATLAS-J2s} (left).
The collective significance of the excess was $4.7\sigma$;  and for the second peak only,  
the  local significance was  $3.0\sigma$.
That peak has a mass and width of $7220 \pm 30 \pm ^{+10}_{-30}$ and $90 \pm 60 ^{+60}_{-30}$~MeV.
Without the 7200~MeV peak, Model~$\beta$ was a single BW, 
shown in Fig.~\ref{fig:ATLAS-J2s} (right), with
a significance of $4.3\sigma$. 
The mass and width were $6960 \pm 50 \pm 30$ and  $510 \pm 170 ^{+110}_{-100}$~MeV---the width 
is much larger than in $J/\psi\,J/\psi$.

This result is substantial evidence for an $J/\psi  \psi(2S)$ excess, but ATLAS did not prefer a particular fit, and  out of caution declined to attribute the excess to any particular origin.

\section{Di-Charmonia: Inferences, Speculations, and Explanations?}
\indent

The \(X(6900)\) structure is firmly established with
ATLAS, CMS, and LHCb all in unison, although parameters depend upon the fit model chosen.
Likewise, the experiments concur on the gross features of the remainder of the mass spectrum, including a dip around 6750~MeV, which favors interference---but different modeling choices  resulted in a somewhat confusing collection of potential structures.
Up to four BW's are needed to describe the $J/\psi\,J/\psi$ spectrum.
The experimental picture may seem muddled, especially for theoretical model builders.


There is  large body of theoretical literature on the $J/\psi\,J/\psi$ structures~\cite{Gutsche:Exotic2024}, but it is not our aim to review this work.
Rather we look at some basic features of the data,  and some simple and gross physics considerations.

\subsection{Some inferences}
\indent

To bring some coherence of the diversity of $J/\psi\,J/\psi$ modeling, we asess  each of the four BW's:

{\bf Threshold BW}: All experiments see evidence that the threshold requires some component to account for the rapid threshold turn-on. 
All experiments were reluctant to assign any particular origin to it: tetraquark, feeddown, rescattering of  dicharmonia
production,\ldots.
A BW seems to provide a viable {\it ad hoc} description.
CMS chose to explicitly relegate it to the background.
What this structure is  stands as an enigma.

$\boldsymbol{X(6600)}$: 
Given its position and broadness there are nagging concerns about contributions from potential feeddown, or that it could be composed of multiple BW's.

$\boldsymbol{X(6900)}$: All experiments observe this structure, but with somewhat varying masses. Its existence is not questioned but its nature is.
Whether tetrquark or  dynamical effect, the $X(6900)$ is accepted as a physics signal.

 $\boldsymbol{X(7100)}$: CMS found a significant ($4.7\sigma$) peak around 7200~MeV in the $J/\psi\,J/\psi$ channel, and hints were seen by ATLAS and LHCb. 
 CMS data favor $X(7100)$ interference, but the hints from ATLAS and LHCb are too weak to discern such an effect.
We provisionally  adopt the $X(7100)$ as a physical structure.

In summary, we adopt the \(X(6600)\), \(X(6900)\), and \(X(7100)\) structures as physical, and remain agnostic about the threshold BW.
We also accept that these three structures likely interfere with each other.

\subsection{Some speculations}
\indent

We will consider some of the broad empirical features of the data and what general lessons they might suggest.
We will on occasion draw on the diquark model of tetraquarks as a  backdrop for these reflections.

One immediate conclusion:  interference among the states, if confirmed, implies they all have the same $J^{PC}$.

\begin{figure}[tb]
    \centering
\begin{minipage}[b]{0.46\textwidth}
    \centering
    \resizebox{0.98\linewidth}{!}{\includegraphics{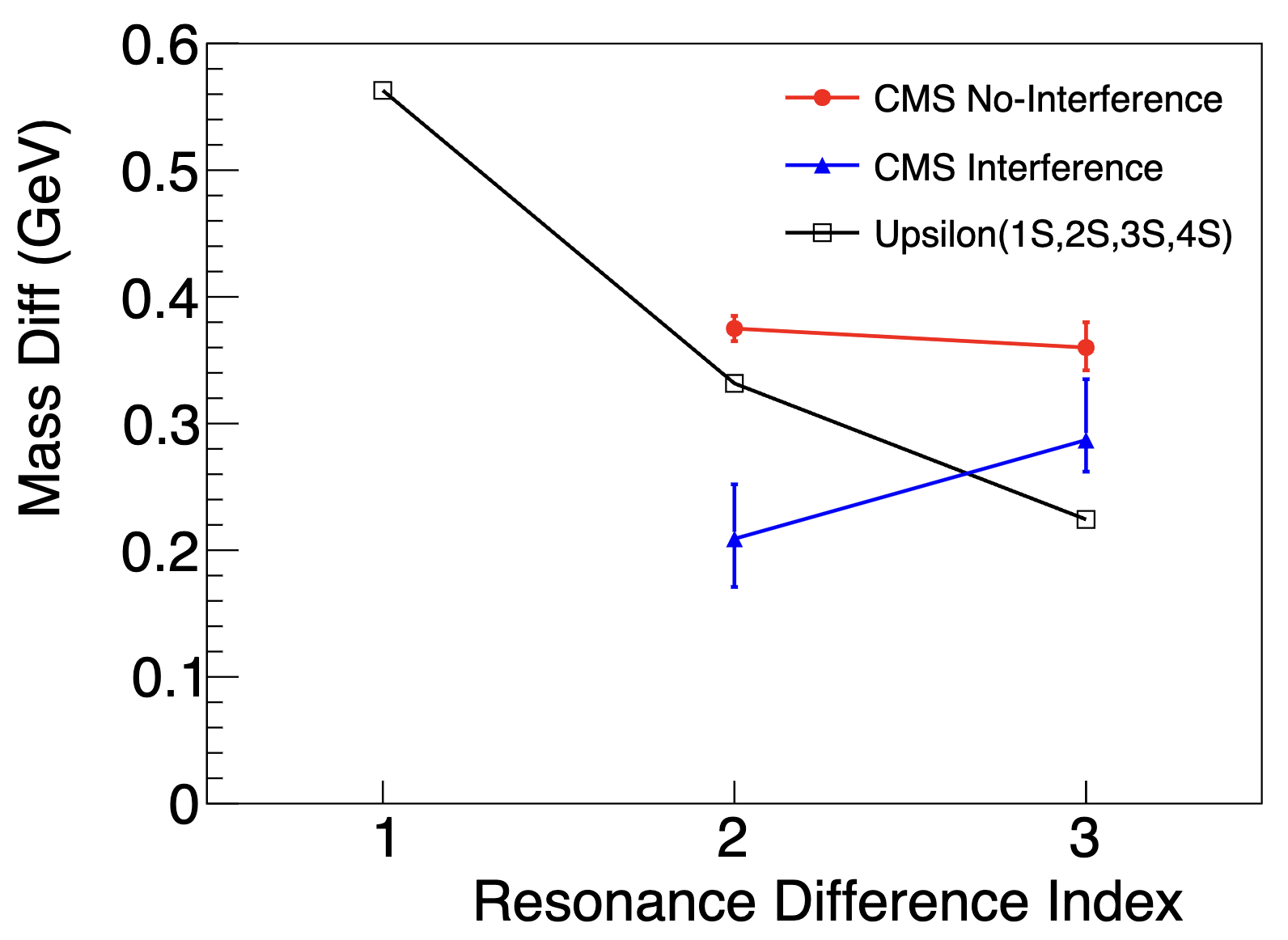}}
    \caption{The mass differences between neighboring states as a function of state index.
    The black line is $\Delta m_i(\Upsilon) = m[\Upsilon(\{i+1\}S)] - m[\Upsilon(iS)]$, where $i$ is their radial index.
    Similarly, the blue (red) line is   $\Delta m_i(X)$ for CMS interference (no-interf.) fits, where the \(X(6600)\), \(X(6900)\), and \(X(7100)\) are assigned indices $i=2$, 3, and 4.
    The plotted uncertainties for $\Delta m_i(X)$ are only statistical, and do not include the \(X(6900)\) contributionl  in order to avoid a correlation in uncertainties between the points.
    \label{fig:JJ-MassDiff}
    }
\end{minipage}
\hfill
\begin{minipage}[b]{0.46\textwidth}
    \centering
    \resizebox{0.98\linewidth}{!}{\includegraphics{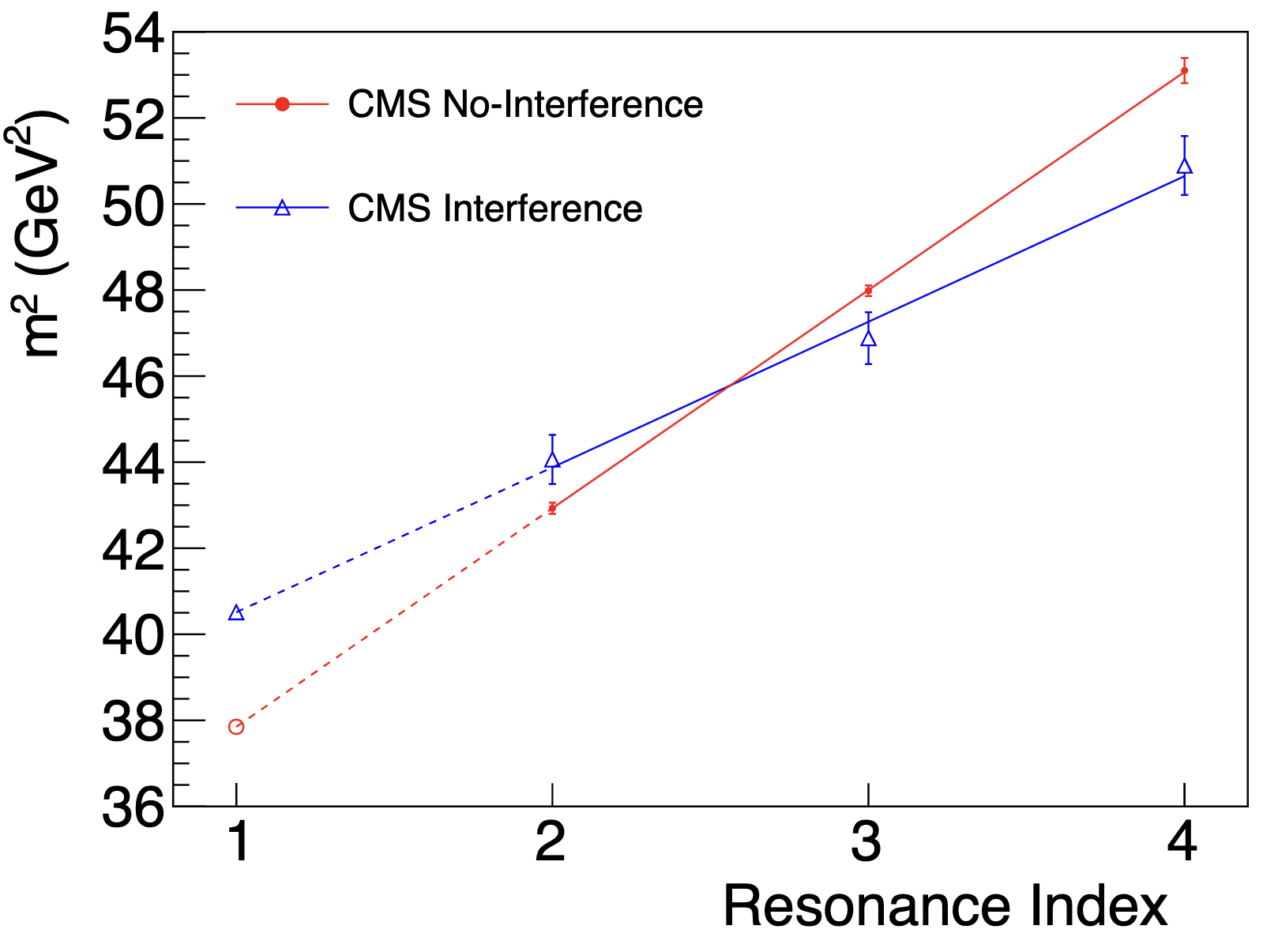}}
    \caption{A Regge-like plot of the square of the CMS \(X(6600)\), \(X(6900)\), and \(X(7100)\) masses as a function of their index $i = 2,3,$ and 4.
    The line is a $\chi^2$ fit of the data.
    The dashed line with open points is an extrapolation for a hypothetical neighboring structure.
    The uncertainties are only statistical.
    \label{fig:JJ-Regge}
    }
\end{minipage}
\end{figure}

Next we consider  mass differences:
they are relatively large ($>$250~MeV), larger than is generally characteristic of spin or orbital mass splittings.
As a reference, we compare CMS's $X$ mass splittings to that of the $\Upsilon$ family in Fig.~\ref{fig:JJ-MassDiff}.
We plot 
$\Delta m_i(\Upsilon) = m[\Upsilon(\{i+1\}S)] - m[\Upsilon{(iS)}]$~\cite{ParticleDataGroup:2024}, with  similar values for
$\Delta m_i(X)$ superimposed.
The indices of the three $X$'s are  shifted to 
$i=3$, 4, and 5 so that the $\Delta m_i$ values of similar magnitude for the $\Upsilon$'s and $X$'s overlap.
The $X$ mass splittings are roughly similar to those for the $2S$--$4S$ $\Upsilon$'s.

A  popular approach to modeling heavy quarkonia  is to use some refinement of the Cornell Potential~\cite{Eichten:1974Cornell1,Eichten:1978Cornell2},  which in its original, and simplest, form is the sum of a Coulomb-like
attraction ($1/r$) for one-gluon exchange and a long-range linear ($r$) confining force.
The two components have the same magnitude around 0.3~fm~\cite{Brambilla:QuarkPot1999}.
The Upsilon sizes for the $1S$, $2S$, and $3S$ have been estimated to be 0.24, 0.51, and 0.73~fm~\cite{Bali:UpsilonSize1997}.
Thus the large $\Upsilon$ mass splitting for $1S$-$2S$ is because the $1S$ state is deeper in the $1/r$ part of the potential, whereas the higher radial excitations are increasingly dominated by the linear potential.
The $X$ splittings, by analogy with the Upsilons, suggest that the triplets are in, or near, the linear regime of the potential.

Cornell-type potentials have also been used to model tetraquarks~\cite{Debastiani:DiquarkSize2017,Yang:DiQuarkSize2020}. 
In some models the diquark structure is exploited to convert the four-body problem into three two-body problems.
Calculations find that the $c$-$c$ separation in a diquark is around $\sim\,$0.4-0.6~fm~\cite{Kiselev:DiquarkSizeBaryon2002,Debastiani:DiquarkSize2017,Lu:DiQuarkSize2020,Yang:DiQuarkSize2020}
.\footnote{Calculations seem to display a curious feature that ground-state diquark separations in the tetraquark, $\sim$0.2-0.4 fm~\cite{Debastiani:DiquarkSize2017,Lu:DiQuarkSize2020}, are {\it smaller} than the $c$-$c$ separation within the diaquark.
This suggests that the approximations that use diquarks to convert the four-body problem into two-body problems are based on an invalid assumption\cite{Debastiani:DiquarkSize2017}, but four-body calculations seem to arrive at similar spacings~\cite{Lu:DiQuarkSize2020}.
But this may suggest that approximating the pair of diquarks as a two-body problem may place them deeper, i.e. more tightly bound, in the Coulombic part of the potential, resulting in lower tetraquark masses.
}
If correct, this prevents the diquark pair from inhabiting much of the Coulombic potential, and would be aligned with  the $X$ states displaying splittings {\it on par} with the higher $\Upsilon$ states.
These crude considerations suggest the $X$ triplet to be a family of radial excitations
(see Ref.~\cite{Lin:4cMass2024} for not-crude considerations); and from a diquark perspective, that the diquarks have a significant separation and reside in, or near, the linear confining part of the potential.


Going a step further, we look at the absolute masses.
A Regge trajectory of the $X$'s as $m^2$ versus a (presumed) radial index ($n_r$)~\cite{Zhu:ReggeJJ2020} is plotted in Fig.~\ref{fig:JJ-Regge} for both CMS's no-interference and interference fits.
The triplet masses in both models are well-described by a linear relationship.
This has several implications:
it supports CMS's identification of these structures as physical states;
the triplet does indeed look like a family; and
it implies any contaminating effects such as feedown, or multiple resonances, have not greatly distorted the masses.
This is especially striking for the no-interference values where uncertainties are much smaller.


The Regge relationship 
$n_r = \beta m^2 + \beta_0$
was used to fit the known charmonium states, and then extrapolate these to Regge parameters for  tetracharm trajectories~\cite{Zhu:ReggeJJ2020}. We compare these to our fits of CMS data:
{
  \begin{center}
    \begin{tabular}{@{}cccc||ccc@{}} 
state & diquark  & $\beta$ (GeV$^{-2}$) & $\beta_0$ &   CMS model & $\beta$ (GeV$^{-2}$) & $\beta_0$\\    \hline
$0^{++}$ & $1^+$  &  $0.206 \pm 0.013$ & $-7.74 \pm 0.56$ & No Interf. &  $0.197 \pm  0.005$ & $-7.46 \pm 0.24$\\
$0^{++}$ & $0^+$  &  $0.212 \pm 0.013$ & $-7.76 \pm 0.57$ & Interf. & $0.289 \pm 0.025$ & $-11.70 \pm 1.23$\\
$2^{++}$ & $1^+$  &  $0.211 \pm 0.013$ & $-8.00 \pm 0.58$ &   &   &  \\
    \end{tabular}
\end{center} }
\noindent 
The slope from CMS's no-interference fit is compatable with the projections, the interference fit disagrees by almost $3\sigma$ (but recall the CMS uncertainties here are only statistical).
The  mass predictions in Ref.~\cite{Zhu:ReggeJJ2020} for the two heaviest states of the  $0^{++}$ ($1^+$ diquark) trajectory  are $6883 \pm 27$ and $7154 \pm 22$~MeV, which agree well with CMS's interference model, but the intermediate state prediction of  $6555 ^{+36}_{-37}$~MeV is about 100~MeV below the interference value---but agrees with the no-interference fit.
CMS included the effect of feedown in the systematic uncertainties of $X(6600)$, but its central value can be biased by this.
A somewhat lower $X(6600)$ mass would actually improve the linear fit for the interference trajectory (Fig.~\ref{fig:JJ-Regge}).
For the  no-interference fit the $X(6600)$  mass has the same problem, but also the $X(7100)$ mass is 130~MeV higher than the projection.
This Regge comparison is not conclusive, 
but both fits
seem well-aligned with a tetraquark interpretation of the triplet.

Another interesting  feature of the triplet  is their natural widths.
The widths from the interference fit significantly shrink as one goes to higher states: 
$\Gamma_{6900}/\Gamma_{6600} = 0.43  ^{+0.26}_{-0.22}$
and
$\Gamma_{7100}/\Gamma_{6900} = 0.51 ^{+0.27}_{-0.20}$,
which are compatible with a common value.
The diquark model offers a simple intuitive possibility: at higher excitations the diquark separation increases, reducing the overlap of diquark wave functions, and thus suppressing decay rates.
$S$-wave states still have significant overlap at the origin, so the large decrease could point to a $P$-wave triplet.
In dramatic contrast,  the  widths from CMS's non-interference fit are consistent with being equal.

The Regge interpretation of the \(X(6900)\) as  $n_r=3$ leaves the   $n_r=1$ state missing.
We can extrapolate our fit to find  this  mass---irrespective of the particular interpretation of the triplet.
Our Regge fit projects a structure at
  $6359 \pm 58$ 
  ($6152 \pm 20$)~MeV 
for the interference (no-interference) fit.
The no-interference mass is below threshold, but perhaps not implausibly so.
ATLAS reported a mass for their threshold-BW in Model~A as $6410 \pm 80 ^{+80}_{-30}$~MeV, consistent with our interference projection.
Other experiments can make a simlar report  from their existing fits.
However, it is a delicate matter to determine precise  resonance parameters with confidence  because of the (unknown) feedown contamination and potential complications from being near threshold.
This is an encouraging first glance at the threshold structure, and perhaps a way to help unravel its mystery.

One may also look in the other direction based on a Regge projection: where might a higher structure lie?
The next structure, if there is one, would be around
$7598 \pm 57$ 
($7951 \pm 28$~MeV) for interference (non-interference) models.
These masses are well above the 
$\Xi_{cc}^+    \Xi_{cc}^-$ and 
$\Xi_{cc}^{++} \Xi_{cc}^{--}$ thresholds ($\sim$7240~MeV~\cite{ParticleDataGroup:2024}), possibly leaving the \(X(7100)\) as the last member of the $J/\psi\,J/\psi$ family.
This is a prime issue for future LHC studies.

\subsubsection{Explanations?}
\indent

The $J/\psi\,J/\psi$ mass spectrum observed at the LHC has proven very fruitful.
The  discovery of the \(X(6900)\)
marks a major milestone in hadronic spectroscopy, and the addition of the \(X(6600)\) and \(X(7100)\) structures introduces critical information to understand the $J/\psi\,J/\psi$ spectrum.
While the remainder of the spectrum has engendered some puzzlement, 
CMS's picture of a triplet of states finds additional support in their orderly pattern of masses, which are in fair agreement with a theoretical projections for the Regge trajectories of $c\bar{c}c\bar{c}$ systems~\cite{Zhu:ReggeJJ2020}.
The Regge character of CMS's triplet lends  support for the three structures havimg the same $J^{PC}$, as implied by the interference model. 
The success of the Regge trajectory model also implies that the mysterious threshold BW makes a plausible candidate for a first radial state.

This tidy tetraquark picture weights against dynamical explanations.
But, it is hard to dismiss the dynamical approach given a degree of  success in their modeling of the $J/\psi\,J/\psi$ spectrum~\cite{Gong:2020bmg,Wang:2020tpt,Huang:NovelDynamics2024CMS},  and the ``coincidence'' of the proximity of exotic structures to pair-production thresholds, e.g.:
 \(X(6600)\) around  
    $\eta_c \chi_{c1}$, $h_c J/\psi$, $\eta_c \psi(2S),\ldots$; 
 \(X(6900)\) around 
    $J/\psi \psi(2S)$, $J/\psi$$\psi(3686)$, $J/\psi$$\psi(3770)$,$\ldots$; 
 \(X(7100)\) around  $\chi_{c2} \chi_{c2}$, $\chi_{c1} \eta_c(2S)$,  $\Xi_{cc} \bar{\Xi}_{cc}$,\ldots. 
 Still,  it seems odd that  purely dynamical threshold  structures would follow a Regge trajectory.


\section{Searches in the Di-Upsilon Spectra}
\indent

The $J/\psi\,J/\psi$ channel opened  the world of all-heavy exotics, but there is much more in this world, in particular $c\bar{c}b\bar{b}$ and $b\bar{b}b\bar{b}$, and non-symmetric  combinations like $c\bar{b}b\bar{b}$.
These are obvious targets for the ever increasing data sets expected from the LHC.
But, in fact, searches for the ultimate all-heavy exotic, $b\bar{b}b\bar{b}$,  have already been formally or informally reported.


\begin{figure}[tb]
    \centering
    \resizebox{0.6\linewidth}{!}{\includegraphics{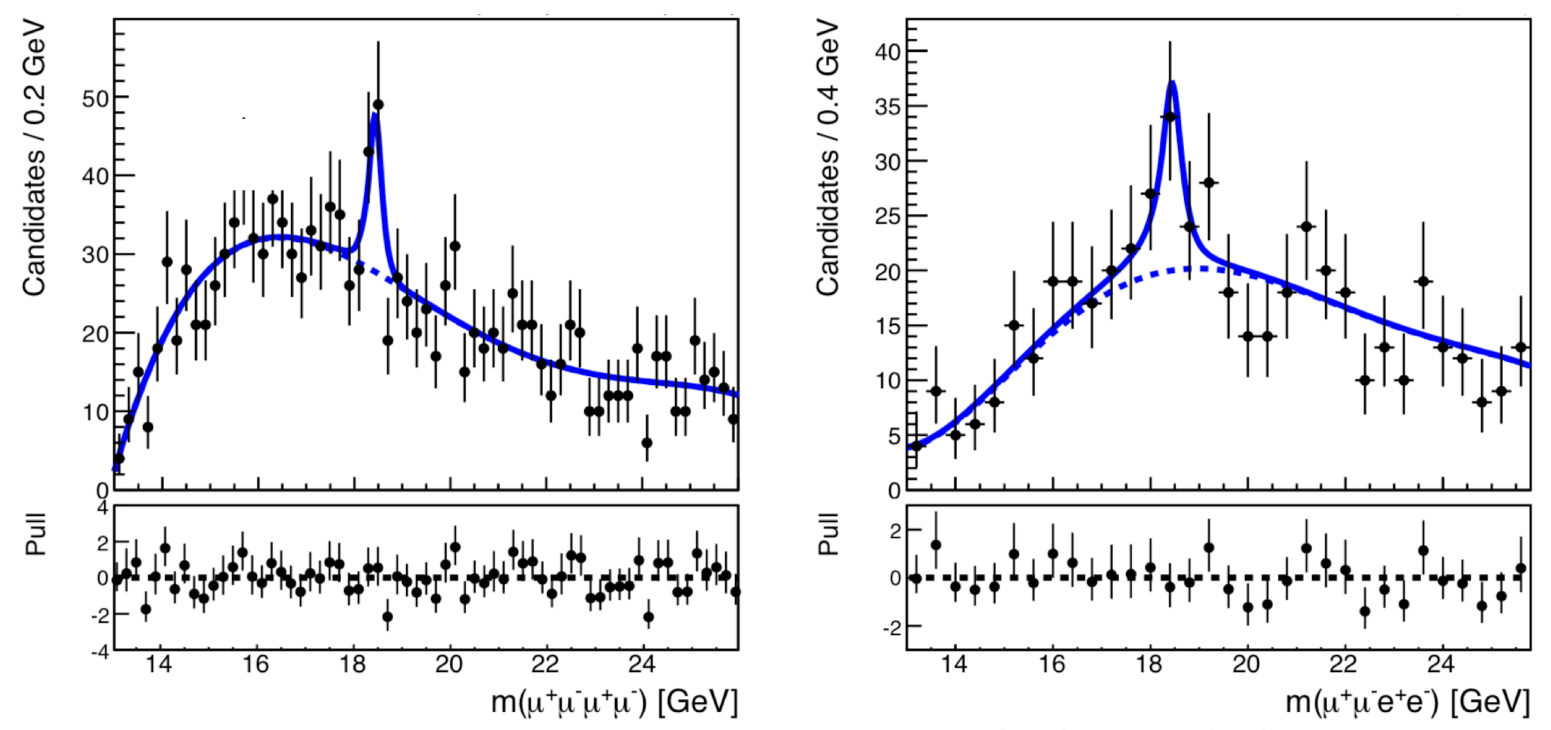}}
    \resizebox{0.38\linewidth}{!}{\includegraphics{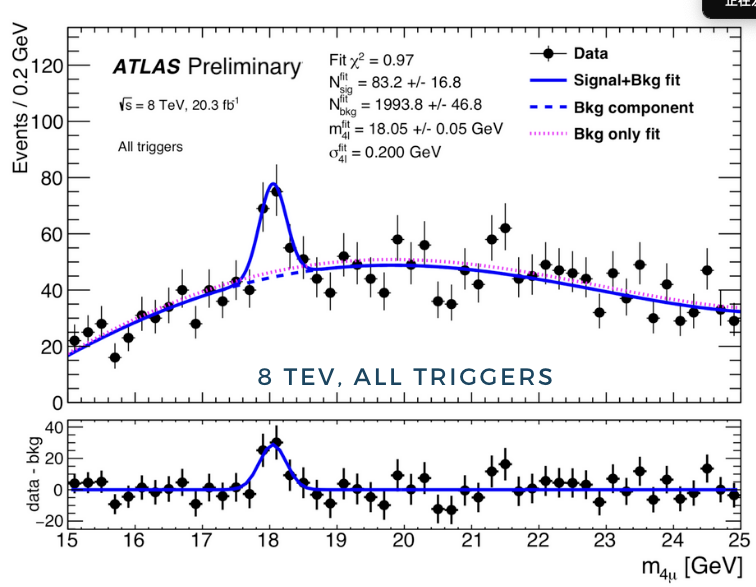}}
    \\

    \caption{The $\Upsilon l^+ l^-$ spectrum for $\mu^+ \mu^- \mu^+ \mu^-$ (left) and $\mu^+ \mu^- e^+ e^-$ (center) channels using CMS 7 and 8 TeV CMS data~\cite{CMS:UpsUps2018Durgut,Durgut:Thesis2018}.
    The ATLAS $\mu^+ \mu^- \mu^+ \mu^-$ distribution for 20~fb$^{-1}$ of 8~TeV data (right)~\cite{ATLAS:18GeV,ATLAS:2023conf,ATLAS:JJ2022conf}.
}
    \label{Fig:18GeV}
\end{figure}

The first observation of $\Upsilon$-pair production was reported by CMS with 21~fb$^{-1}$ of 8~TeV data
in 2017~\cite{CMS:2016liw}.
Only a year later,  a conference report of a Ph.D. thesis described
structure in the $\Upsilon l^+ l^-$ channel using the $\mu^+ \mu^- \mu^+ \mu^-$ ($\sim\! 25$~fb$^{-1}$) and $\mu^+ \mu^- e^+ e^-$ ($\sim\! 20$~fb$^{-1}$) modes with  CMS data at 7 and 8~TeV~\cite{CMS:UpsUps2018Durgut, Durgut:Thesis2018,Yi:2018fxo}.
A preliminary signal was seen in each channel as shown in Fig.~\ref{Fig:18GeV} (left and center).
With a $4\mu$ signal of $44\pm 13$ candidates ($3.9\sigma$ local significance) the  mass was
$18.4 \pm  0.1  \pm 0.2$~GeV. 
In the $2e2\mu$ mode, $35 \pm 13$ candidates ($3.2\sigma$) measured a mass of
$18.5 \pm 0.2 \pm 0.2$~GeV.
The two modes agree, and when combined yield a mass of $18.4 \pm 0.1 \pm 0.2$~GeV, for a local significance of $4.9\sigma$, and globally $3.6\sigma$.

These are fairly significant results, but they were challenged by another $\Upsilon l^+ l^-$ search by LHCb with 6~fb$^{-1}$ of data at 7, 8, and 13~TeV in 2018, where nothing was seen in a 17.5-20.0~GeV window~\cite{LHCb:2018uwm}.
%
In due course CMS published a result in this channel with 36~fb$^{-1}$ of 13~TeV data from 2016---and now also found no signal~\cite{CMS:2020qwa}.
This was not much more data than CMS analyzed before, but it was at almost twice the energy.

Energy is not a very plausible excuse for a signal to vanish,\footnote{
While an energy difference is not a very likely explanation, higher energies come with larger cross sections, typically increased instantaneous luminosities, and higher pileup background,
which usually forces triggers and selections to be  tightened. 
It is conceivable that an $\Upsilon l^+ l^-$ signal was lost through the increased contamination of pile-up events and the tightened trigger selection.
This might be unravelled if the signal resurfaces, and experimenters wished to do some reconstructive archaeology.
} 
so the tale becomes even more bizarre with a 2023 conference report by ATLAS where they seemed to  confirm CMS's 18~GeV peak---just before unconfirming it~\cite{ATLAS:18GeV,ATLAS:2023conf,ATLAS:JJ2022conf}!
With 20~fb$^{-1}$ of 8~TeV data ATLAS reported an $18.05 \pm 0.05$~GeV peak in the $4\mu$ channel with a $5.4\sigma$ local significance (Fig.~\ref{Fig:18GeV}, right).
But in 13~TeV data from 2015-17 (52~fb$^{-1}$) the ``signal'' was an insignificant $1.9\sigma$ with the mass was fixed to 18.05~GeV; and then with data from 2018 (59~fb$^{-1}$, with a different trigger) they were unable to confirm any structure: there was no upward ``fluctuation'' seen at all!
Weird\ldots but an object lesson on the insidious nature of vanishing signals.

The LHC was not alone.
A surprising entry in the  all-bottom tetraquark sweepstakes comes from heavy-ion collisions.
In 2019 a preprint by the A$_N$DY Collaboration at RHIC claimed a peak  with a dijet signature ~\cite{ANDY:2019bfn}.
Jet mass resconstructions have been the domain of very heavy particles, like 
$W$, $Z$, and top----and not for  hadron spectroscopy.
A$_N$DY looked at dijets triggered in Cu$+$Au collisions at $\sqrt{s_{NN}}=200$~GeV recorded in 2012\footnote{
The data was collected for a calorimeter test, and as a result the luminosity was not measured.}
and found a peak in the forward dijet mass of
$18.12 \pm 0.15 \pm 0.6$~GeV, with significance over $9\sigma$.
The peak is several GeV wide, and thus presumably would be a collective effect of multiple (many?) hypothetical tetraquarks.
Unfortunately this paper remains unpublished, and there seems to be no further reports.

Finally, it should also be noted that some theoretical expectations, such as lattice QCD \cite{Hughes:18GevQCD}, predict that there are no all-$b$ tetraquark states below the $\eta_b \eta_b$ threshold, i.e. 18.76 GeV.
%
In any case, the LHC is churning out more data, and a compelling demonstration of all-bottom structures might not be too far off?

\section{Summary and Conclusions}
\indent

The quark model has become an irrevocable, but not immutable, core of the Standard Model of particle physics.
Originally perceived as a crazy and bizarre idea in 1964,  quarks became real in the 1970's, and the model was followed by a relentless march of successes.
At the pinnacle of its success at the turn of the century the model was confronted by a new twist: particles not conforming to the conventional quark configurations---exotic hadrons.
This was not entirely a surprise, as some investigators had been toiling at the periphery chasing nature's occasional teasers of exotics, but also stumbling over a number of false alarms.
The discovery of the $X(3872)$ meson in 2003 ignited an explosion of exotic hadrons.
The oddities included pentaquarks  and charged charmonium-like states ($Z^+$'s).
Despite this bounty of states, there continues much debate as to what is their exact natures: molecular, diquark, hybrid, or dynamical threshold effects.

The old lesson of simplicity with heaviness learned from the $\psi$ and $\Upsilon$ families pointed down the road to seek systems composed entirely of heavy quarks.
It took six years to find the first $s\bar{s}q\bar{q}$ candidate in $J/\psi\phi$, but not without several years of controversy.
Then, finally in 2020, LHCb found the $X(6900)$ structure in $J/\psi\,J/\psi$---the ideal candidate for an all-charm tetraquark.
The $J/\psi\,J/\psi$ spectrum was more complex than this single structure, and CMS reported observing the  $X(6600)$ structure in excess of $5\sigma$, opening the door to study of the {\it spectrum} of tetracharm states. 
A bonus was their $X(7100)$, which had a substantial significance even if
below the $5\sigma$ benchmark.
It is  by virtue of having  a triplet that all of our speculations and explanatory musings, right or wrong, become possible. 

But having reached this pot of gold, we are confronted with the question: exactly what is their nature?

We reviewed the experimental results from ATLAS, CMS, and LHCb on the $J/\psi\,J/\psi$ spectrum, and sifted through the various fit models to  winnow all the hypothetical Breit-Wigner ``resonances'' down to a basic set of physical candidates.
The data  paints a picture of at least
three structures, 
$X(6600)$, $X(6900)$,  and $X(7100)$, and possibly (probably?) a fourth at threshold.
The triplet states appear to  mutually interfere---thereby implying the same $J^{PC}$'s.
The mass splittings are fairly large ($>$250~MeV), suggesting a family of radial excitations.
This picture is endorsed by the $X$ masses conforming to a Regge trajectory of radial excitations.
The pattern of significantly shrinking natural widths in the triplet prompts a simple appeal to a diquark model.

This picture is by no means conclusive.
While resorting to interference has been universal and is strongly motivated, the data is not statistically conclusive and the possibilities of non-interference modelling has not been fully exhausted (e.g. multi-resonance peaks).
It is noteworthy that the no-interference fit of CMS data also conforms to a Regge trajectory---but this too implies common quantum numbers.
Issues of interference, and the possibility of further states, will be addressed with the ongoing collection of larger data samples at the LHC, as well as new types of measurements will become possible, such as production cross sections, new modes, branching fractions, and spin-parity determinations.
More data also means all-heavy tetraquarks containg bottom quarks, with  signatures like $J/\psi \Upsilon, \Upsilon \Upsilon,\Upsilon \Upsilon^*$  ($\Upsilon^*$ is off-shell $\Upsilon$ decaying to dileptons), may also be in reach soon.
One can also expect searches in new directions and modes,
such as Belle's look for structure in  $\eta_c J/\psi$ directly from $e^+ e^-$ annhilation~~\cite{Belle:EtacJPsi2023}.

The $X(3872)$ sparked a revolution in spectroscopy which as has been echoing to this day.
The coming years promise exciting new chapters in the physics of exotic hadrons, and the exploration of QCD in this new domain.

\newpage
\section{Acknowledgments}
\indent

We would like to thank the following colleagues for their careful reviews and helpful suggestions: 
Liupan An, Evelina Bouhova, Kai-feng Chen, Xin Chen, Dmitri Denisov, Andrei Gritsan, Yuanning Gao, Zhen Hu, Shan Jin, Georgios Karathanasis, Dmytro Kovalskyi,  Haibo Li, Xinchou Lou, Gautier Hamel de Monchenault, Stephen Olsen, Sergey Polikarpov, Darren Price, Chiara Rovelli,  Chengping Shen, Lenny Spiegel, Karim Trabelsi, Adam Wolfgang, Changzheng Yuan, Jingqing Zhang, Liming Zhang, Bing Zhou. However, all opinions and comments expressed, and any error committed, are solely responsibility of the authors.

\bibliographystyle{unsrt}
\bibliography{refer}
\end{document}